%% file: main.tex
\documentclass[aps,prx,twocolumn,superscriptaddress,amsmath]{revtex4-2}
\usepackage[hidelinks]{hyperref}
\usepackage[separate-uncertainty=true,binary-units]{siunitx}
\usepackage{graphicx}
\usepackage{tikz}
\usepackage{glossaries}
\usepackage{ifthen}
\usepackage[noabbrev,capitalise]{cleveref}
\usepackage{import}
\usepackage{rotating}
\usepackage[T1]{fontenc}

\usetikzlibrary{
	calc,
	shapes,
	positioning,
	backgrounds,
	decorations.pathreplacing,
	fit,
	external,
	fadings}
\usetikzlibrary{external}
\pgfdeclarelayer{background}
\pgfdeclarelayer{foreground}
\pgfsetlayers{background,main,foreground}

\definecolor{nicegray}{HTML}{555555}
\definecolor{nicered}{HTML}{AF5A50}
\definecolor{niceblue}{HTML}{005B82}
\definecolor{nicegreen}{HTML}{7D966E}
\definecolor{niceyellow}{HTML}{D7AA50}

\crefname{figure}{FIG.}{FIG.}
\crefname{table}{TABLE}{TABLE}
\crefname{equation}{Eq.}{Eq.}
\crefname{supp}{SI}{SI}

\newacronym{ei}{E-I}{excitatory-inhibitory}
\newacronym{lif}{LIF}{leaky integrate-and-fire}
\newacronym{asic}{ASIC}{application-specific integrated circuit}
\newacronym{ppu}{PPU}{plasticity processing unit}
\newacronym{simd}{SIMD}{single instruction multiple data}
\newacronym{hmm}{HMM}{Hidden Markov Model}
\newacronym{psp}{PSP}{post-synaptic potential}
\newacronym{epsp}{EPSP}{excitatory post-synaptic potential}

\begin{document}

\title{Autocorrelations from emergent bistability in homeostatic spiking neural networks on neuromorphic hardware}
\author{Benjamin Cramer}
\author{Markus Kreft}
\author{Sebastian Billaudelle}
\author{Vitali Karasenko}
\author{Aron Leibfried}
\author{Eric Müller}
\author{Philipp Spilger}
\author{Johannes Weis}
\author{Johannes Schemmel}
\affiliation{Kirchhoff-Institute for Physics, Im Neuenheimer Feld 227, Heidelberg University, Germany}
\author{Miguel A. Mu\~noz}
\affiliation{Departamento de Electromagnetismo y Física de la Materia e Instituto Carlos I de Física Teórica y Computacional, Universidad de Granada, E-18071 Granada, Spain}
\author{Viola Priesemann}
\affiliation{Max Planck Institute for Dynamics and Self-Organization, Am Fa{\ss}berg 17, 37077 G\"ottingen, Germany}
\affiliation{Institute for the Dynamics of Complex Systems, University of G\"ottingen, Friedrich-Hund-Platz 1, 37077 G\"ottingen, Germany}
\author{Johannes Zierenberg}
\affiliation{Max Planck Institute for Dynamics and Self-Organization, Am Fa{\ss}berg 17, 37077 G\"ottingen, Germany}

\date{\today}

\input{content/abstract.tex}
\pacs{}
\keywords{bistability}

\maketitle

\glsresetall

\input{content/introduction.tex}
\input{content/methods.tex}
\input{content/results.tex}
\input{content/discussion.tex}
\input{content/acknowledgements.tex}

\appendix
\input{content/appendix.tex}

\bibliography{benjamin_bistable.bib}

\setcounter{section}{0}
\renewcommand{\thefigure}{S\arabic{figure}}
\setcounter{figure}{0}
\renewcommand{\thetable}{S\arabic{table}}
\setcounter{table}{0}
\renewcommand\appendixname{Supplementary Information}

\clearpage
\onecolumngrid
\section*{Supplementary Information}
\input{content/si.tex}

\end{document}

%% file: content/abstract.tex
\begin{abstract}
A unique feature of neuromorphic computing is that memory is an implicit part of processing through traces of past information in the system's collective dynamics.
The extent of memory about past inputs is commonly quantified by the autocorrelation time of collective dynamics.
Based on past experimental evidence, a potential explanation for the underlying autocorrelations are close-to-critical fluctuations.
Here, we show for self-organized networks of excitatory and inhibitory \acrlong{lif} neurons that autocorrelations can originate from emergent bistability upon reducing external input strength.
We identify the bistability as a fluctuation-induced stochastic switching between metastable active and quiescent states in the vicinity of a non-equilibrium phase transition.
This bistability occurs for networks with fixed heterogeneous weights as a consequence of homeostatic self-organization during development.
Specifically, in our experiments on neuromorphic hardware and in computer simulations, the emergent bistability gives rise to autocorrelation times exceeding $\SI{500}{\milli\second}$ despite single-neuron timescales of only $\SI{20}{\milli\second}$.
Our results provide the first verification of biologically compatible autocorrelation times in networks of \acrlong{lif} neurons, which here are not generated by close-to-critical fluctuations but by emergent bistability in homeostatically regulated networks.
Our results thereby constitute a new, complementary mechanism for emergent autocorrelations in networks of spiking neurons, with implications for biological and artificial networks, and introduces the general paradigm of fluctuation-induced bistability for driven systems with absorbing states.
\end{abstract}

%% file: content/introduction.tex
\section{\label{sec:introduction}Introduction}

Neuromorphic computing covers a variety of brain-inspired computers, devices, and models that function fundamentally different to common von-Neumann architectures~\cite{schuman_survey_2017-1, furber_large-scale_2016}.
For instance, one can mimic information processing in the brain by \emph{emulating} the dynamics of neuron membrane potentials and synaptic currents in electronic circuits~\cite{schemmel_modeling_2007, schemmel_wafer-scale_2010, friedmann_demonstrating_2017, moradi_scalable_2018, benjamin_neurogrid_2014}, exploiting our brains' low energy cost in energy-efficient hardware.
In contrast to common computers, the neuromorphic hardware (e.g. the neural network) is responsible for both processing and memory at the same time.
How exactly processing and memory are represented is highly hardware specific, calling for hand-in-hand development of hardware and corresponding algorithms, which can be guided by modern artificial intelligence and neuroscience likewise~\cite{markovic_physics_2020}.
For spiking neural networks, an active question is how past information can be ``stored'' in the collective neural dynamics.

Empirical work from neurophysiology and neuroimaging suggests that neural circuits implement process memory as active traces of past information~\cite{hasson_hierarchical_2015}.
One measure of the extend over which information can be accumulated is the processing timescale, which has been related to the autocorrelation time of collective neural dynamics~\cite{murray_hierarchy_2014, hasson_hierarchical_2015}.
It was found that the autocorrelation time is related to the hierarchical anatomy of cortex~\cite{murray_hierarchy_2014, hasson_hierarchical_2015, raut_hierarchical_2020, spitmaan_multiple_2020, gao_neuronal_2020, siegle_survey_2021}: timescales are shorter in primary sensory regions and longer in higher-order cortical regions.
In fact, empirical observations of autocorrelation times in neural recordings range from $\mathcal{O}(\SI{10}{ms})$ to $\mathcal{O}(\SI{1}{s})$~\cite{murray_hierarchy_2014, hasson_hierarchical_2015, raut_hierarchical_2020, spitmaan_multiple_2020, gao_neuronal_2020, siegle_survey_2021, wasmuht_intrinsic_2018, cavanagh_reconciling_2018, wilting_operating_2018, wilting_between_2019}.
This seems to contradict prior theoretical predictions for networks of \gls{ei} \gls{lif} neurons, where early theories and models of networks of \gls{lif} neurons in an \gls{ei} balanced state \cite{vreeswijk_chaos_1996,renart_asynchronous_2010} predict almost vanishing mean correlations, though more recent reassessments found conditions under which much larger correlations can appear~\cite{rosenbaum_spatial_2017, mastrogiuseppe_intrinsically-generated_2017, baker_correlated_2019}, see also Ref.~\cite{latham_correlations_2017, dahmen_global_2022} for an overview.
Focusing on temporal correlations, recent developments in dynamic mean-field theory~\cite{ostojic_two_2014, mastrogiuseppe_intrinsically-generated_2017, van_meegen_microscopic_2021} reveal parameter ranges with emergent autocorrelation times larger, but still on the order of the characteristic time of the membrane potential, typically $\mathcal{O}(\SI{10}{ms})$, which is still far below the ones observed experimentally.

This raises the question whether the emergence of large autocorrelation times requires additional mechanisms of self-organization.
A generic mechanism of neural self-organisation is \emph{homeostatic plasticity}, a negative feedback that adapts local neural properties to achieve a stable firing rate~\cite{turrigiano_activity-dependent_1998, turrigiano_homeostatic_2004, turrigiano_homeostatic_2012}.
For homeostatically regulated excitable systems, one can prove analytically that lowering the input strength induces an increase in the recurrent coupling and hence in the autocorrelation time through close-to-critical fluctuations~\cite{zierenberg_homeostatic_2018}.
These predictions are consistent with experiments on monocular deprivation~\cite{ma_cortical_2019} and with the hierarchy of timescales~\cite{murray_hierarchy_2014, hasson_hierarchical_2015, raut_hierarchical_2020, spitmaan_multiple_2020, gao_neuronal_2020, siegle_survey_2021}, and have been applied as a guiding principle to tune neuromorphic hardware for optimal task performance~\cite{cramer_control_2020}.
However, it so far remains unknown whether close-to-critical fluctuations are responsible for emergent autocorrelations in homeostatic regulated \gls{ei}  networks of \gls{lif} neurons, or whether there could be alternative mechanisms.

It is thus necessary to (i) verify that  large autocorrelation times can emerge in networks of \gls{ei} \gls{lif} neurons and to (ii) develop a theoretical understanding of their origin.
Here, we address these open questions using the experimental setup of a neuromorphic chip with homeostatic plasticity during development.
We verify that reducing the external input strength induces increasing autocorrelation times that can be more than 20 times larger than the decay time of the membrane potential of individual units.
We complement our size-constrained experiments with a numerical finite-size scaling analysis that reveals progressively larger autocorrelation times for increasing system sizes.
Surprisingly, we find that in our case autocorrelations are not generated by close-to-critical fluctuations~\cite{zierenberg_homeostatic_2018}, but originate from an emergent bistability in the population firing rate.
To explain this bistability, we derive a mean-field theory for driven excitable systems that reveals a fluctuation-induced switching between a metastable active phase and a quiescent phase, reminiscent of so-called \emph{up and down} states in brain networks~\cite{wilson_up_2008,stern_spontaneous_1997,cossart_attractor_2003,hidalgo_stochastic_2012}.
We finish with a discussion of how emergent bistability can affect biological and artificial neural networks, as well as other finite systems with an absorbing-to-active transition that are driven by external noise.

%% file: content/methods.tex
\section{\label{sec:methods}Model and Methods}

\paragraph{Neural network.}
As a minimal model of biological spiking neurons, we consider a recurrent network of $N=512$ (unless otherwise stated) \gls{lif} neurons coupled to an input layer consisting of $N/2$ Poisson neurons (\cref{fig:chip}A).
Each \gls{lif} neuron integrates spikes from, on average, $K^\mathrm{rec}=100$ recurrent neurons of the network and from, on average, $K^\mathrm{ext}$ external neurons of the input layer.
The \emph{physical} connection between neuron $i$ and $j$ is randomly realized, $c_{ij}=\{-1,0,1\}$, and further weighted by an integer-value coupling weight $w_{ij}\in[0, 63]$.
Neurons can be either excitatory or inhibitory, which is reflected in the sign of $c_{ij}$ for a given neuron $j$ for all outgoing coupling synapses. Also, in analogy with cortical networks~\cite{douglas_inhibition_2009}, \SI{20}{\percent} of the neurons in both network and input layer are inhibitory. While the recurrent neurons are \gls{lif} neurons, input neurons generate spikes independently as a Poisson process with rate $\nu^\mathrm{ext}=\SI{10}{\hertz}$, which amounts to an average input rate per recurrent neuron of $h=\nu^\mathrm{ext}K^\mathrm{ext}$.

The dynamics of a recurrent \gls{lif} neuron $i$ is modelled by a leaky membrane potential $u_i(t)$ given by
\begin{equation}
	\tau^\mathrm{m}_i\dot{u}_i(t) = u_i^\mathrm{leak}-u_i(t) + R_i I_i(t)\, , \label{eq:membrane}
\end{equation}
where $\tau^\mathrm{m}_i = C_i^\mathrm{m} R_i$ is the membrane time constant with the membrane capacitance $C_i^\mathrm{m}$ as well as the resistance $R_i$, and $u_i^\mathrm{leak}$ is the leak potential.
Similarly, $I_i(t)$ denotes a leaky synaptic current which is described by
\begin{align}
    \tau^\mathrm{s}_i\dot{I}_i(t) = -I_i(t) &+ \gamma_i\sum_j c_{ij}w_{ij}\sum_k\delta(t-t_j^k-\tau^d)\, , \label{eq:synapse}
\end{align}
where $\tau^\mathrm{s}_i$ is the synaptic time constant, $\gamma_i$ is a scale factor, $w_{ij}$ are dimensionless coupling weights between neurons $i$ and $j$ (which covers recurrent and external pre-synaptic neurons), and $\sum_k\delta(t-t_j^k-\tau^\mathrm{d})$ is the spike train of neuron $j$ that arrives at neuron $i$ with past spike times $t_j^k$ and synaptic time delay $\tau^\mathrm{d}$.
Spikes are generated once a neuron's membrane potential crosses a threshold, i.e., $u_i(t)>u_i^\mathrm{thres}$, after which the membrane potential is reset to $u_i^\mathrm{reset}$ where it remains for the duration of the refractory period $\tau^\mathrm{ref}$.
Explicit parameters were motivated by neurophysiology but subject to experimental constraints further specified in the next section listed (cf. \cref{tab:parameters}).

While external coupling weights are fixed to $w_{ij}^\mathrm{in} = 17$, recurrent couplings are subject to homeostatic plasticity during development to regulate the single-neuron firing rates around a target rate $\nu^\ast=\SI{10}{\hertz}$.
More specifically, we implement homeostatic plasticity as an iterative, stochastic update of all realized ($c_{ij}\neq 0$) recurrent weights $w^\mathrm{rec}_{ij}$.
Each iteration consists of a $\SI{5}{\second}$ time window for which we record the firing rate $\nu_i$ of each individual neuron.
In between iterations, we update each recurrent weight $w^\mathrm{rec}_{ij}$ with a (small) probability $p$ by and amount
\begin{equation}
	\Delta w_{ij} = \lambda(\nu^\ast-\nu_j),
	\label{eq:rule}
\end{equation}
which depends only on the local information of the post-synaptic neuron $j$ and where $\lambda p$ sets the time scale of the homeostasis.
While we here chose a small probability $p$ to overcome artifacts from the limited precision of $w_{ij}$ (see \cref{sec:hom_parametrization} for the effect of $p$ and $\lambda$), we obtain similar results when instead updating each weight by $\Delta w_{ij}$ plus integer rounding noise (see \cref{sec:hom_comparison}).
To preserve the effective sign of excitatory and inhibitory weights, $w_{ij}$ are restricted to positive values and saturate at zero.
Besides this, the proposed simple update scheme does not distinguish between excitatory and inhibitory couplings.
After the application of $\Delta w_{ij}$, the network dynamics are evolved for about \SI{1}{\second} in order to allow the network activity to equilibrate before assessing $\nu_i$ for the next update.
Importantly, we only employ homeostatic plasticity during the development stage of our experiment but afterwards keep all weights fixed during the acquisition of the spike data on which the evaluation of autocorrelations is performed.

\paragraph{Neuromorphic chip.}
We emulate \gls{ei} networks of $N=512$ \gls{lif} neurons on the mixed-signal neuromorphic system BrainScaleS-2~\cite{friedmann_demonstrating_2017,schemmel_accelerated_2022,pehle_brainscales-2_2022} (\cref{fig:chip}B).
The term \textit{emulation} is used to clearly distinguish between this physical implementation, where each observable has a measurable counterpart on the neuromorphic chip, and standard software \textit{simulations} on conventional hardware (see below).
In particular, neurons are implemented as electrical circuits that emulate the \gls{lif} dynamics outlined in \cref{eq:membrane} in a time-continuous and parallel manner (see \cref{sec:appendix_hardware}).
Due to the analog implementation, time constants are determined by the electrical components on the substrate and are rendered approximately a factor \num{1000} times faster than the ones of their biological archetype.
Within this paper, all referenced time scales are converted to the equivalent biological time domain unless otherwise stated.

The system consists of an array of $256 \times 512$ physically implemented current-based synapses that supports near arbitrary topologies.
Their dynamics emulate leaky currents as detailed in \cref{eq:synapse} and feature coupling strengths $w_{ij}$ with a precision of \SI{6}{\bit} (i.e., $64$ discrete values).

Homeostatic plasticity is implemented on-chip by a specialized, freely-programmable processor, the \gls{ppu}~\cite{friedmann_demonstrating_2017}.
The latter is able to update the synaptic weights of \SI{128}{} synapses in parallel.
To implement updates according to \cref{eq:rule}, we draw on dedicated circuits within each neuron that count the number of emitted spikes.

The system comes with specialized accelerators for the drawing of random numbers~\citep{schemmel_accelerated_2020}.
These facilitate an on-chip generation of Poisson distributed input spikes as well as the efficient implementation of the stochastic homeostatic regulation without additional communication bottlenecks.
The only remaining communication with the host system consists of the transfer of instructions for configuring the BrainScaleS-2 system at the beginning and the readout of the result at the end of an experiment~\citep{muller_extending_2020,muller_scalable_2022}, making the hardware implementation very fast.

The neuromorphic chip is subject to parameter variations both in space and time:
first, the analog implementation causes temporal noise within the model dynamics and second, the production process necessarily leads to small variations across electrical components.
The latter variations can, however, be mitigated by exploiting the configurability of the BrainScaleS-2 system by resorting to calibration routines~\citep{muller_scalable_2022}, thereby reducing the parameter spread across neurons (see \cref{sec:parametrization}).
The remaining variability of parameters can be quantified by their mean and standard deviation (\cref{tab:parameters}).

\paragraph{Computer Simulation.}
For comparison and finite-size scaling analysis, we use additional computer simulations where we employ the Python simulation package Brian2~\cite{goodman_brian_2009}.
This package generates from the differential equations \cref{eq:membrane,eq:synapse} a discrete-time Euler integration scheme together with full control over all system parameters.
We use these simulations to (i) crossvalidate the results from the neuromorphic chip (see \cref{sec:validation}) and to (ii) analyze how changing system sizes, $N=\{256,512,768,1024\}$, beyond the hardware-limiting constraints affect our conclusions.
The integration time step is set to $\Delta t = \SI{50}{\micro\second}$ to approach the time continuous nature of the BrainScaleS-2 system.
To closely mimic the emulated networks, we draw the individual neuron parameters from Gaussian distributions specified by the measured parameter variability of the neuromorphic chip (\cref{tab:parameters}).
In addition, independent temporal noise with standard deviation $\sigma \sqrt{2*\tau^\mathrm{m}_i}$, with $\sigma = \SI{2}{\milli\volt}$, is added to \cref{eq:membrane}.
To ensure reproducibility, the code has been made freely available~\cite{noauthor_benjamincramerneuromorphic-bistability_nodate}.

\paragraph{Observables.}
The main observables we consider are derived from the \textit{instantaneous population firing rate} $\nu(t)$, defined as the number of network spikes within a time bin $\Delta t$
\begin{equation}
	\nu(t)=\frac{1}{N\Delta t}\sum_{i=1}^N \sum_{k=1}^{S_i}\int_t^{t+\Delta t}\delta(t-t_i^k),
\end{equation}
where $t_i^k$ are the spike times of neuron $i$, $S_i$ the number of spikes emitted by neuron $i$, and $\Delta t=\SI{5}{\milli\second}$.

From a time series $\nu(t)$, we calculate the stationary \textit{autorrelation function}
\begin{equation}\label{eq:correlation}
	C(t^\prime) = \frac{\text{Cov}[\nu(t+t^\prime)\nu(t)]}{\text{Var}[\nu(t)]},
\end{equation}
where $t^\prime$ are multiples of $\Delta t$.
From the decay of the autocorrelation function it is possible to derive the time scale(s) of temporal correlations.
Specifically, if the autocorrelation function can be described by a single exponential decay, $C(t^\prime)=e^{-t^\prime/\tau_\mathrm{AC}}$, one can extract a single autocorrelation time $\tau_\mathrm{AC}$ by fit routines.

To estimate statistical errors, we average over \num{50} independent experiments.

%% file: content/results.tex
\section{\label{sec:results}Results}

\subsection{Homeostatically regulated neuromorphic hardware compensates lack of external input by strengthening recurrent connections}

\begin{figure*}[ht]
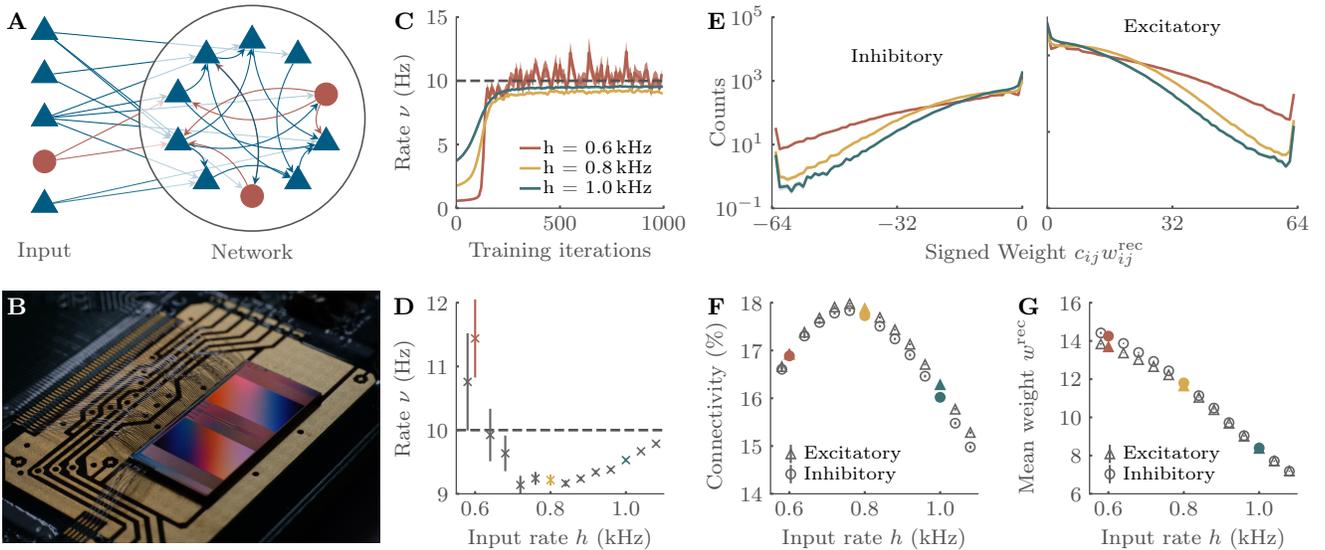

	\centering
	\begin{tikzpicture}
		\node[anchor=north west,inner sep=0pt] (b) at (0,-3.8) {\includegraphics[width=.58\columnwidth]{images/chip.JPG}};
		\node[anchor=north west,inner sep=0pt] (a) at (0.1,0.0) {\input{content/network}};
		\node[anchor=north west,inner sep=0pt] (c) at (5.15,0.0) {\input{build/figures/rate_evolution.pgf}};
		\node[anchor=north west,inner sep=0pt] (d) at (5.15,-3.8) {\input{build/figures/final_rate.pgf}};
		\node[anchor=north west,inner sep=0pt] (e) at (9.3,0.0) {\input{build/figures/weight_histogram.pgf}};
		\node[anchor=north west,inner sep=0pt] (f) at (9.3,-3.8) {\input{build/figures/connectivity.pgf}};
		\node[anchor=north west,inner sep=0pt] (g) at (13.45,-3.8) {\input{build/figures/average_weight.pgf}};
		\node at ($(a.north west) + (0.1,-0.2)$) {\textbf{A}};
		\node[text=white] at ($(b.north west) + (0.2,-0.2)$) {\textbf{B}};
		\node at ($(c.north west) + (0.2,-0.2)$) {\textbf{C}};
		\node at ($(d.north west) + (0.2,-0.2)$) {\textbf{D}};
		\node at ($(e.north west) + (0.2,-0.2)$) {\textbf{E}};
		\node at ($(f.north west) + (0.2,-0.2)$) {\textbf{F}};
		\node at ($(g.north west) + (0.2,-0.2)$) {\textbf{G}};
	\end{tikzpicture}

	\caption{%
		\textbf{Reducing input strength to homeostatically regulated networks of \gls{ei} \gls{lif} neurons strengthens recurrent connections.}
		\textbf{(A)} Illustration of a random network topology with \SI{80}{\percent} excitatory (blue triangles) and \SI{20}{\percent} inhibitory (red circles) neurons.
		\textbf{(B)} Image of the BrainScaleS-2 neuromorphic chip.
		Image taken from \citep{muller_extending_2020}.
		\textbf{(C)} Homeostatic plasticity regulates the population rate $\nu$ close to a target value (dashed line).
		\textbf{(D)} For a broad range of external input rates, $\nu$ approximates the target rate.
		\textbf{(E)} The stochastic homeostatic regulation leads to heterogeneous weight distributions for both, inhibitory and excitatory synapses.
		The counts of excitatory weights exceed the inhibitory ones by a factor of four due to the imposed \gls{ei} ratio.
        	\textbf{(F)} The effective connectivity, defined as the percentage of non-zero recurrent synapses ($c_{ij}w^\mathrm{rec}_{ij}\neq 0$), does not saturate at its maximum for decreasing input strengths.
		\textbf{(G)} However, the mean weight increases to compensate for a reduction of input.
	}
	\label{fig:chip}
\end{figure*}

To begin, we verify that the experimental setup --- the neuromorphic chip with homeostatic regulation during development --- reaches a stationary dynamical state with a firing rates sufficiently close to the target rate.
Starting from the initial condition of zero recurrent weights ($w^\mathrm{rec}_{ij}=0$), we observe for our chosen parameters a transient relaxation behavior that reaches a stationary firing rate after about \SI{200}{} update iterations, independent of the external input rate $h$ (\cref{fig:chip}C).
Note that for this representation, the firing rate is evaluated over an interval of $T=\SI{100}{\second}$ between iterations, and further averaged over \num{50} network realizations.
One can see that for larger $h$ (blue curve) the relaxation is smoother than for lower $h$ (red curve).
The stationary firing rates are close to the target rate $\nu^\ast=\SI{10}{\hertz}$ (\cref{fig:chip}D), however, there is a systematic $h$-dependence that presumably originates from the firing rate being a non-linear function of the mean coupling, $\nu(\langle w\rangle)$, as observed in mean-field calculations of \gls{ei} networks~\cite{mastrogiuseppe_intrinsically-generated_2017}.
Since we find consistent results for reference computer simulations (see \cref{sec:validation}), we conclude that the experimental setup reliably self-organizes into a stationary dynamics state with neuron firing rates close to the target rate.

We next investigate how homeostatically regulated \gls{ei} networks compensate a reduction of external input with a strengthening of recurrent connections (\cref{fig:chip}E-G).
In particular, we find that the histograms of both inhibitory as well as excitatory recurrent coupling weights become broader with decreasing $h$, indicating strong heterogeneity (\cref{fig:chip}E).
It is noteworthy that the limited range of weight values is an effect of the \SI{6}{\bit} integer weights on the neuromorphic hardware.
Interestingly, the effective connectivity -- the fraction of potentially non-zero weights that remains zero -- does not reach its maximum theoretical value of $K/N = 100 / 512 \approx \SI{20}{\percent}$ (\cref{fig:chip}F).
Instead, it even decreases for low $h$, which is likely a consequence of the strong variability of rates between iterations (cf. \cref{fig:chip}C) that results in large weight changes for the given plasticity rule and does not affect our main conclusions (see \cref{sec:hom_comparison} for a milder plasticity rule).

More important is the observation that mean coupling weights $w^\mathrm{rec}$ increase almost linearly with decreasing input rate (\cref{fig:chip}G).
A fit of the form
\begin{equation}\label{eq:w-h}
    \langle w^\mathrm{rec}\rangle(h) = \alpha - \beta h,
\end{equation}
where $\langle.\rangle$ stands for average across synaptic connections over either excitatory and inhibitory populations, yields $\alpha\approx\SI{22.75}{}$ and $\beta\approx\SI{14.23}{}$ for excitatory or $\alpha\approx\SI{26.1}{}$ and $\beta\approx\SI{-16.7}{}$ for inhibitory weights.
Hence, a reduction in input rate clearly strengthens the recurrent connections in homeostatically regulated \gls{ei} networks consistent with the theoretical consideration that the loss of external input needs to be compensated by recurrent activity generation in order to maintain a constant firing rate~\cite{zierenberg_homeostatic_2018}.

\begin{figure*}[ht]
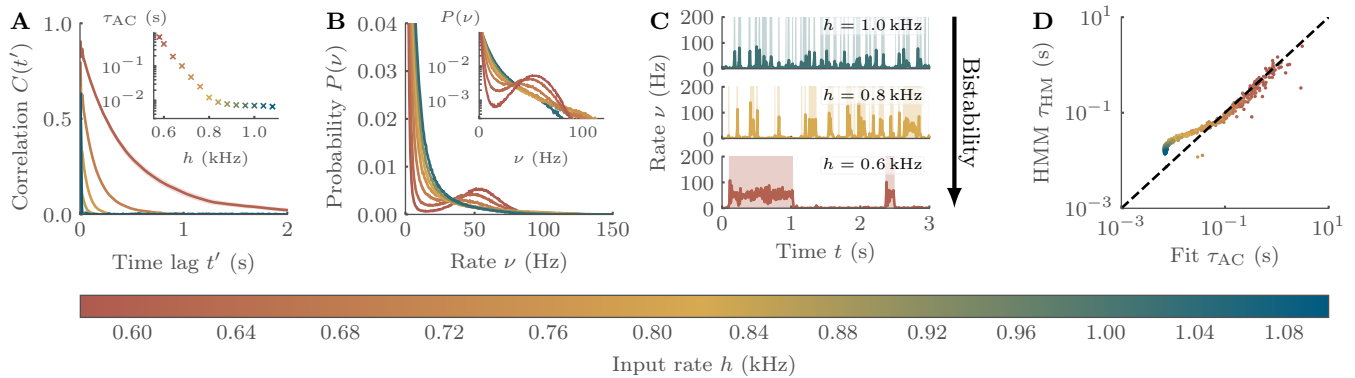

	\centering
	\begin{tikzpicture}
		\node[anchor=north west,inner sep=0pt] (x) at (0.94,-3.8) {\import{build/figures/}{colorbar.pgf}};
		\node[anchor=north west,inner sep=0pt] (a) at (0,0) {\input{build/figures/ac.pgf}};
		\node[anchor=north west,inner sep=0pt] (b) at (4.2,0) {\input{build/figures/activity_distribution.pgf}};
		\node[anchor=north west,inner sep=0pt] (c) at (8.5,0) {\input{build/figures/activity.pgf}};
		\node[anchor=north west,inner sep=0pt] (d) at (13.6,0) {\input{build/figures/timescale_comparison.pgf}};
		\node at ($(a.north west) + (0.2,-0.2)$) {\textbf{A}};
		\node at ($(b.north west) + (0.2,-0.2)$) {\textbf{B}};
		\node at ($(c.north west) + (0.2,-0.2)$) {\textbf{C}};
		\node at ($(d.north west) + (0.2,-0.2)$) {\textbf{D}};

		\draw[-latex,ultra thick] ($(c.north east) + (0.2,-0.15)$) -- ($(c.south east) + (0.2,0.8)$) node[midway,xshift=0.2cm,rotate=-90] {Bistability};
	\end{tikzpicture}
	\caption{%
		\textbf{Reducing input strength increases autocorrelation of network rate through emergent bistability.}
		\textbf{(A)} For low input rates $h$, the population activity exhibits exponentially shaped autocorrelations $C(t')$ with autocorrelation times $\tau_\mathrm{AC}$ significantly exceeding the largest single neuron timescale.
		\textbf{(B)} In this regime, the distribution $P(\nu)$ of the population rate $\nu$ shows a bimodal trend.
		\textbf{(C)} The associated phases of high and low $\nu$ can be fitted by a two state \acrshort{hmm}.
		\textbf{(D)} Based on the transition rates of this \acrshort{hmm}, an equivalent timescale $\tau_\mathrm{HM}$ can be estimated which coincides with $\tau_\mathrm{AC}$ for low $h$.
	}
	\label{fig:time_scale}
\end{figure*}

In addition to supporting general theoretical arguments, our setup allows us to investigate how our homeostatic self-organization affects the interplay between excitatory and inhibitory neurons.
In fact, it is quite surprising that the mean coupling weights for excitatory and inhibitory weights are so similar (\cref{fig:chip}G), i.e., $\langle w^\mathrm{rec}_\mathrm{inh} \rangle \approx \langle w^\mathrm{rec}_\mathrm{exc}\rangle$, given that each neuron receives four times more input from excitatory than from inhibitory neurons.
Naively, this implies strong excitation dominance in contrast to the expected inhibition dominance required for asynchronous irregular activity~\cite{vreeswijk_chaos_1996,brunel_dynamics_2000} to reproduce experimental single-neuron statistics~\cite{burns_spontaneous_1976, softky_highly_1993, stevens_input_1998, stein_neuronal_2005}.
This outcome can, however, be explained by our symmetric plasticity rule that does not distinguish between excitatory and inhibitory synapses and thereby fosters solutions with $\langle w^\mathrm{rec}_\mathrm{inh} \rangle \approx \langle w^\mathrm{rec}_\mathrm{exc}\rangle$, which turn out to be feasible transition points between high- and low-firing phases of our small networks (see \cref{sec:phase_diagrams}).
We further show in \cref{sec:phase_diagrams} that the corresponding transition is between regular and irregular dynamics and that reducing $h$ makes this transition sharper, implying that homeostatic plasticity regulates \gls{ei} networks closer to the regular-to-irregular transition for decreasing external input rate.

\subsection{Homeostatically regulated neuromorphic hardware with low external input generates large autocorrelation times through emergent bistability}

Next, we verify the theoretical prediction~\cite{zierenberg_homeostatic_2018} that a homeostatically regulated system exhibits an increased autocorrelation to compensate for a decreasing external input (\cref{fig:time_scale}).
For this, we consider a network after homeostatic development with fixed weights and evaluate the autocorrelation function of $\nu(t)$ over an interval of $T=\SI{100}{\second}$.
Indeed, the autocorrelation functions, $C(t^\prime)$, show increasingly long tails with decreasing input rate $h$ (\cref{fig:time_scale}A).
Moreover, they are well described by exponential decays, $C(t^\prime)=e^{-t^\prime/\tau_\mathrm{AC}}$, with increasing autocorrelation times $\tau_\mathrm{AC}$ for decreasing $h$ (\cref{fig:time_scale}A inset).
While this general trend has been reported for much smaller neuromorphic systems before~\cite{cramer_control_2020}, here, we clearly observe the emergence of two distinct regimes.
For $h>\SI{0.8}{\kilo\hertz}$, we find autocorrelation times to saturate with increasing $h$, suggesting that the uncorrelated Poisson input successfully decorrelates already weakly correlated activity, giving rise to an \textit{input-driven regime}.
In contrast, for $h<\SI{0.8}{\kilo\hertz}$, we find an apparent divergence of $\tau_\mathrm{AC}$ with decreasing $h$, such that this regime is characterized by dominant recurrent activation compensating for the lacking input, which results in increasing autocorrelation times for decreasing $h$, giving rise to a \textit{recurrent-driven regime}.

Surprisingly, we observe that the autocorrelations originate from a bistable population rate (\cref{fig:time_scale}B-D).
Specifically, distributions of $\nu(t)$ change from unimodal for high input strengths to bimodal for lower input strengths (\cref{fig:time_scale}B), the latter suggesting that the population rate starts to alternate between two distinct states.
Indeed, close inspection of the time evolution of $\nu(t)$ reveals that for decreasing input strength the population rate switches between a low-rate state and a high-rate state (\cref{fig:time_scale}C), resembling up-and-down states in cortical networks~\cite{wilson_up_2008,stern_spontaneous_1997,cossart_attractor_2003,hidalgo_stochastic_2012}.
This is fundamentally different from the a-priori expected close-to-critical fluctuations~\cite{zierenberg_homeostatic_2018}, which would lead to scale-free avalanches~\cite{beggs_neuronal_2003} for small $h$ that we do not observe (see \cref{sec:avalanches}).

Instead, the network dynamics are better described as a Markov jump process between states of high- and low-firing rates~\cite{ibe_3_2013}.
The population rate can be well approximated by a two-state \gls{hmm}~\cite{rabiner_introduction_1986}.
From the \gls{hmm}, we obtain a transition matrix $p_{ij}^\mathrm{t}$ that allows us to calculate the autocorrelation time of the Markov jump process as $\tau_\mathrm{HM}=-\Delta t/ \log{(\lambda_2)}$ with $\lambda_2$ being the second largest eigenvalue of $p_{ij}^\mathrm{t}$.
Indeed, the autocorrelation time of the \gls{hmm} correlates with the autocorrelation time measured from the population rate for small input strengths, where the population rate becomes bistable (\cref{fig:time_scale}D).
We thus conclude that the emergent bistability is the underlying mechanism of the large autocorrelation times observed in the population dynamics of homeostatically regulated \gls{ei} networks of \gls{lif} neurons.

\subsection{Computer simulations reveal increasing dynamical barrier of emergent bistability with system size}

Since the neuromorphic hardware only supports networks with up to $N=512$ \gls{lif} neurons, we use computer simulations to verify the experimental results for increasing network sizes.
In brief, we parametrize the simulations to match the experimental setup and use the Brian2 Python package to solve the model (for details see \cref{sec:methods}).
Indeed, we can reproduce the experimental results with our software implementation: We observe comparable bistable activity with similar autocorrelation functions (see \cref{sec:validation}).
However, while computer simulations in principle allow us to study any system size, they are much less efficient than the neuromorphic emulation.
It is worth noting that for our application, i.e., homeostatically regulating a network of $N=512$ \gls{lif} neurons for \SI{6000}{\second}, the computer simulation on an Intel Xeon E5-2630v4 (roughly $\SI{100 000}{\second}$ at about $\SI{50}{\watt}$) takes $\mathcal{O}(10^4)$ more time and $\mathcal{O}(10^7)$ more energy compared to the corresponding emulation on BrainScaleS-2 (about $\SI{6}{\second}$ at a power budget of \SI{100}{\milli\watt}).

\begin{figure}[t]
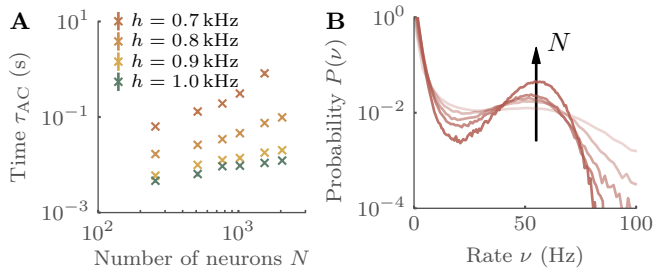

	\centering
	\begin{tikzpicture}
		\node[anchor=north west,inner sep=0pt] (a) at (0,0) {\input{build/figures/simulation_ac.pgf}};
		\node[anchor=north west,inner sep=0pt] (b) at (4.2,0) {\input{build/figures/simulation_activity_distribution.pgf}};
		\node at ($(a.north west) + (0.2,-0.2)$) {\textbf{A}};
		\node at ($(b.north west) + (0.2,-0.2)$) {\textbf{B}};
	\end{tikzpicture}
	\caption{%
		\textbf{Finite-size scaling of homeostatically regulated \gls{ei} networks with \gls{lif} neurons from computer simulations.}
		\textbf{(A)} Autocorrelation time $\tau_\mathrm{AC}$ as a function of system size $N$ for different external input rates $h$.
		One can see a clear increase in $\tau_\mathrm{AC}$ with $N$ for $h<\SI{1}{\kilo\hertz}$, while $\tau_\mathrm{AC}$ seems to saturate for $h>\SI{1}{\kilo\hertz}$.
		\textbf{(B)} Distributions of the population firing rate in windows of $\SI{5}{\milli\second}$ for $h=\SI{0.7}{\kilo\hertz}$ show that bimodal shape is maintained for increasing $N$.
		The barrier in between high- and low-firing states grows with $N$.
    	}
	\label{fig:sim}
\end{figure}

Having established that the computer simulation reproduces the experimental results, we can study how the measured autocorrelation time $\tau_\mathrm{AC}$ depends on the network size $N$ (\cref{fig:sim}A).
Due to the large computational efforts, we focus on three representative input strengths: a low input strength ($h=\SI{0.7}{\kilo\hertz}$) where we observe bistable activity in the experiment, two medium input strengths ($h=\SI{0.8}{\kilo\hertz}$ and $h=\SI{0.7}{\kilo\hertz}$) near the onset of bistability, and a high input strength ($h=\SI{1.0}{\kilo\hertz}$) where the network does not exhibit bistability.
Only for $h=\SI{0.7}{\kilo\hertz}$, we observe a clear increase in autocorrelation time with system size that exceeds $\SI{1}{s}$ for the largest $N$.
Our numerical results further corroborate the classification into two distinct regimes: A recurrent-driven regime for low input strength with large emergent autocorrelations and the input-driven regime for high input strength with vanishing autocorrelations.

To further investigate the origin of the emergent autocorrelations, we study the shape of the probability distribution of local population rates $\nu(t)$ as a function of network size (\cref{fig:sim}B).
We observe that for low input strength, the bimodal distribution becomes more pronounced with increasing suppression of intermediate population-rate values.
One can relate the suppression of intermediate rates to a \textit{dynamical barrier} by interpreting the time course of the instantaneous population rates as a trajectory of the dynamical system in the potential $V(\nu) = -\log P(\nu)$.
This barrier would be analogous to the activation energy in an Arrhenius-type equation, i.e., $r\propto e^{-\Delta V/T}$, such that for a given level of fluctuation $T$ the rate $r$ to transition between low- and high-firing-rate regimes is lowered for increasing barriers $\Delta V$.
Since the height of this dynamical barrier increases with $N$, this explains the increasing autocorrelation time with system size.

\begin{figure}[t]
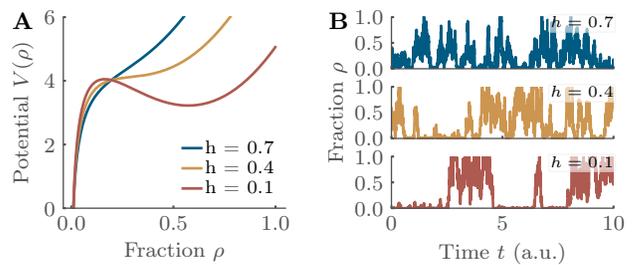

	\begin{tikzpicture}
		\node[anchor=north west,inner sep=0pt] (a) at (0,0)  {\input{build/figures/theory_potential.pgf}};
		\node[anchor=north west,inner sep=0pt] (b) at (4.2,0){\input{build/figures/theory_activity.pgf}};
		\node at ($(a.north west) + (0.2,-0.2)$) {\textbf{A}};
		\node at ($(b.north west) + (0.2,-0.2)$) {\textbf{B}};
	\end{tikzpicture}
	\caption{%
        \textbf{Mean-field theory of emergent bistability upon reducing input to homeostatically regulated recurrent network.}
        Our mean-field theory describes the temporal evolution of the fraction of active neurons, $\rho$, with meta-stable solutions given by the minimum of the potential, \cref{eq:potential}.
        \textbf{(A)} For suitable parameters ($\tau_\mathrm{MF}=10$, $\alpha=30$, $\beta=15$, $b=25$, $\sigma=50$, $N=512$), the potential exhibits a single minimum for large $h$ but two minima for small $h$.
        \textbf{(B)} Numeric evaluation of the corresponding stochastic mean-field equation ($\Delta t=10^{-7}$) shows fluctuating dynamics for large $h$ and emergent bistability for low $h$.
        }
	\label{fig:theory}
\end{figure}

\subsection{Mean-field theory of emergent bistability from fluctuation-induced switching between metastable active and quiescent states}

To qualitatively explain how bistability can emerge in a homeostatically regulated recurrent network, we construct a mean-field theory that describes the time evolution of the fraction of active neurons at a given time $t$, $\rho(t)$, which is proportional to the population rate $\nu(t)$.
Let us consider a general mean-field ansatz
\begin{align}
	\dot{\rho}(t) = &-\tau_\mathrm{MF}\rho(t) \nonumber \\
		&+h  (1-\rho(t)) + \left( 1-\rho(t) \right) \left[ \omega_1\rho(t) + \omega_2\rho^2(t)+\dots\right]\, ,
\end{align}
where the first term describes the spontaneous decay of activity in the absence of inputs with some characteristic time scale
$\tau_\mathrm{MF}$, the term proportional to $h$ represents external input that can only activate inactive neurons, and the last terms represent the gain function that describes recurrent activations, expanded in power-series of the activity.
Here, the coefficients of expansion ($\omega_1$, $\omega_2$, ...) are an effective representation of the full coupling matrix $w^\mathrm{rec}_{ij}$ (with $\omega_1$ proportional to the mean synaptic strength) and the factor $\left(1-\rho(t)\right)$ in the last two terms accounts for activity saturation.
The mean-field equation can be rewritten in a more compact form by grouping-up terms with different powers of the activity,
\begin{equation}\label{eq:MF_compact}
    \dot{\rho}(t) = h - a\rho(t) -b\rho^2(t) + \dots \, ,
\end{equation}
where $a=\tau_\mathrm{MF}+h-\omega_1$ and $b=\omega_1-\omega_2>0$ to ensure stability.
It is important to notice that this mean-field equation assumes infinitely large network sizes, $N\to\infty$, for which additional noise terms vanish.

To describe finite networks one needs to introduce an additional stochastic term to the mean-field equation \cref{eq:MF_compact} that accounts for demographic fluctuations.
Demographic fluctuations are characteristic of systems with an absorbing or quiescent state~\cite{henkel_absorbing_2008}, where fluctuations of the total number of active units around some mean $N\rho(t)$ are expected to have a standard deviation that scales with $\sqrt{N\rho(t)}$ as a consequence of the central-limit theorem.
For the fraction of active nodes in a finite network, we then obtain to leading order in system size
\begin{equation}\label{eq:MF_noise}
    \dot{\rho}(t) = h - a\rho(t) -b\rho^2(t) + \sqrt{\rho(t)/N}\eta(t)\, ,
\end{equation}
where $\eta(t)$ is Gaussian white noise with zero mean and variance $\sigma^2$.
This (Ito) Langevin equation can be expressed as a Fokker Planck equation, with the steady-state solution~\cite{munoz_nature_1998} (see also SI)
\begin{equation}
    P(\rho) = \mathcal{N}\exp\left\{-\frac{2N}{\sigma^2}V(\rho)\right\}\, ,
\end{equation}
a normalization constant $\mathcal{N}$, and the potential
\begin{align}\label{eq:potential}
    V(\rho) = \left(\frac{\sigma^2}{2N} - h\right)\ln\rho + a\rho + \frac{b}{2}\rho^2\, .
\end{align}
This potential $V(\rho)$ can either have a single (formally diverging) minimum at $\rho=0$ (unimodal activity distribution), or it can have two local minima (bistable activity distribution).
Specifically, we can find from the condition $\rho\,dV/d\rho=(\frac{\sigma^2}{2N}-h) + a\rho +b\rho^2=0$ extrema at $\rho_\pm = \left(-a\pm\sqrt{a^2-4b(\frac{\sigma^2}{2N}-h)}\right)/2b$ such that a bistable solution occurs when
    $a^2-4b(\frac{\sigma^2}{2N}-h)>0$.
With the additional conditions for a positive density, i.e., $\rho>0$, as well as a positive slope at $\rho=0$ (cf.~\cref{fig:sim}), i.e., $\rho^2\frac{d^2V}{d\rho^2}(0)=(\frac{\sigma^2}{2N}-h)>0$, we expect to observe bistable dynamics for
\begin{equation}\label{eq:condition-bistable}
    a<-2\sqrt{b\left(\frac{\sigma^2}{2N}-h\right)}<0.
\end{equation}

To incorporate the homeostatic regulation effectively, we recall that in our experiments we found an anticorrelation between $\langle w \rangle$ and $h$ (\cref{fig:chip}G).
The most dominant effect of this anticorrelation, \cref{eq:w-h}, is to alter the coefficient $a=\tau_\mathrm{MF} + h - \omega_1 \approx \tau_\mathrm{MF}  - \alpha + h(1+\beta)$ in our mean-field equation.
Then the condition for a bistable solution, \cref{eq:condition-bistable}, becomes $\tau_\mathrm{MF}-\alpha +h(1+\beta) < -2\sqrt{b(\frac{\sigma^2}{2N}-h)}$ and we find that ---for suitable parameters--- lowering $h$ does indeed induce a transition from a unimodal to a bimodal potential (\cref{fig:theory}).

Numerically evaluating the mean-field model with input-dependent recurrent interactions reproduces bistable population activity (\cref{fig:theory}B).
The numerical integration of \cref{eq:MF_noise} is straightforward~\cite{dornic_integration_2005}, but needs special care to avoid leaving the domain due to numerical imprecisions (see Appendix~\ref{sec:appendix_meanfield}).
The resulting trajectories show typical demographic fluctuations for higher inputs and bistable activity for lower input.
Since the involved parameters are not easily related in an explicit way to the experiment, this theoretical results is a qualitative explanation of the observed effect and all parameters are in arbitrary units.

Our mean-field theory implies that emergent bistable population activity can be rationalized as a fluctuation-induced switching between a metastable active and a quiescent phase.
For a system with an absorbing to active non-equilibrium phase transition for vanishing input, we find that finite-size fluctuations are responsible for a metastable active state (high rate) and external fluctuations lead to a metastable quiescent state (low rate).
To transit from one state to another, the system needs to overcome a dynamical barrier, where the transition from high-to-low rate requires demographic noise, whereas the transition from low-to-high rate requires external noise.

%% file: content/network.tex
\tikzset{%
	inh/.style = {%
		draw=nicered,%
		fill=nicered,
		circle,
		inner sep=0pt,
		minimum width=0.3cm,
	},%
	exc/.style = {%
		draw=niceblue,%
		fill=niceblue,
		regular polygon,
		regular polygon sides=3,
		minimum width=0.4cm,
		inner sep=0pt,
	},%
}%
\pgfmathsetseed{42}%

\begin{tikzpicture}[
		x=2.3cm,
		y=2.3cm,
		>=stealth,
		line width=1.0\pgflinewidth,
		anchor=center
        ]

	\useasboundingbox (-1.4,0.65) rectangle (0.7,-0.82);

	\foreach \b in {0,1,...,4}{
		\ifthenelse{\b=3}{
			\node[inh] (background \b) at (-1.2,0.5-0.25*\b) {};
		}{
			\node[exc] (background \b) at (-1.2,0.5-0.25*\b) {};
		}
	}

	\pgfmathsetmacro{\N}{9}
	\pgfmathsetmacro{\r}{0.45}
	\coordinate (center) at (0,0);
	\foreach \n in {0,1,...,\N}{
		\ifthenelse{\n=2 \OR \n=5}{
			\node[inh] (hidden \n) at ($({\r*sin(\n*360/(\N+1))},{\r*cos(\n*360/(\N+1))})+(center)$) {};
		}{
			\node[exc] (hidden \n) at ($({\r*sin(\n*360/(\N+1))},{\r*cos(\n*360/(\N+1))})+(center)$) {};
		}
	}

        \begin{scope}[on background layer]
			\begin{scope}
			\path[scope fading=south] (-0.8,-0.1) rectangle ++(1.6,-2.0);
			\foreach \h in {0,1,...,\N}{
				\foreach \i in {0,1,...,4}{
    	        			\pgfmathparse{rnd}
    	        			\pgfmathsetmacro{\foobar}{\pgfmathresult}
    	        			\ifthenelse{\lengthtest{\foobar pt<0.2 pt}}{
						\ifthenelse{\i=3}{
							\draw[-stealth,nicered] (background \i) -- (hidden \h);
						}{
							\draw[-stealth,niceblue] (background \i) -- (hidden \h);
						}
					}{}
				}
			}
			\end{scope}
			
			\fill[white,opacity=0.7] (0,0) circle (1.5cm);
			\draw[nicegray,semithick] (0,0) circle (1.5cm);

			\foreach \i in {0,1,...,\N}{
				\foreach \j in {0,1,...,\N}{
					\ifthenelse{\equal{\j}{0}}{
					}{
						\pgfmathparse{rnd}
						\pgfmathsetmacro{\foobar}{\pgfmathresult}
						\ifthenelse{\lengthtest{\foobar pt<0.2 pt}}{
							\pgfmathsetmacro{\bend}{ifthenelse(\j <= \N / 2 + 1,"bend right","bend left"))}
							\pgfmathsetmacro{\target}{int(Mod(\i + \j, \N + 1))}
							\ifthenelse{\i=2 \OR \i=5}{
								\draw[-stealth,nicered] (hidden \i) to[\bend] (hidden \target);
							}{
								\draw[-stealth,niceblue] (hidden \i) to[\bend] (hidden \target);
							}
						}{}
				}
				}
			}
		\end{scope}

	\coordinate (label) at (0.0,-0.77);
	\node[nicegray,anchor=center] at (label -| hidden 0) {\footnotesize Network};
	\node[nicegray,anchor=center] at (label -| background 0) {\footnotesize Input};
\end{tikzpicture}

%% file: content/discussion.tex
\section{\label{sec:discussion}Discussion}

In summary, we showed that homeostatically regulated networks of \gls{ei} \gls{lif} neurons can self-organize towards a dynamical regime of emergent fluctuation-induced bistability with large autocorrelations.
This supports the principal understanding that emerging autocorrelations can compensate a lack of input through recurrent activation to preserve the local neuron firing rate~\cite{zierenberg_homeostatic_2018}.

Here, we identify the existence of bistable population activity as the mechanism of emergent autocorrelations, which complements recent observations of emergent autocorrelations in networks regulated by spike-timing-dependent plasticity~\cite{cramer_control_2020} or after training spiking recurrent neural networks on working memory tasks~\cite{kim_strong_2021}.
This bistability is a result of finite-size fluctuations that have a non-vanishing probability to cross a dynamical barrier between meta-stable collective states, which may remind of free-energy barriers in classical systems such as nucleation~\cite{feder_homogeneous_1966, kashchiev_nucleation_2000, zierenberg_canonical_2017}.
We showed numerically that the dynamical barrier increases with system size, but that dynamical bistability persists for finite networks, albeit with increasing transition times.
Our results thereby provide a new mechanism to generate large autocorrelations in finite neural networks, which are an important ingredient for bio-inspired computing.

Importantly, the fluctuation-induced bistable population activity we observe does not require an active adaptation mechanism:
Our results on bistability were recorded after turning off homeostatic plasticity.
While plasticity was necessary to shape the weight distribution, it is not relevant for the stochastic switching between metastable-active (high rate) and metastable-absorbing (low-rate) states.
The basic mechanism observed here, namely a dynamical barrier that separates two metastable fixed points, is consistent with previous observations of perturbation-induced state switching in spiking neural networks~\cite{brunel_effects_2001, renart_mean-driven_2006, tartaglia_bistability_2017}.
Here, we do not require additional, strong external perturbations, because we can control the height of the dynamical barrier and thereby observe the state switching induced by the available, weak external input during finite-time recordings.
Importantly, the fluctuation-induced bistability is different from adaptation-based mechanisms, such as adaptation currents~\cite{parga_network_2007}, or depletion of synaptic utility~\cite{millman_self-organized_2010, bonachela_self-organization_2010}.

It is interesting to discuss more in depth the connection with the mechanism of self-organized bistability~\cite{di_santo_self-organized_2016, buendia_feedback_2020} which has recently shown to be relevant for collective brain dynamics \cite{buendia_self-organized_2020}.
This is a mechanism akin to self-organized criticality (SOC)~\cite{bak_self-organized_1988, buendia_feedback_2020}, but in which the system self-organizes by means of a feedback-loop between the level of activity (overall firing rate) and the control parameter (the mean synaptic weight) to the edge of a discontinuous phase transition with bistability, rather than to a critical point in systems with SOC.
In the present case, there is a similar feedback loop during development.
But rather than acting on a global control parameter, this feedback acts differentially for each synaptic weight, thereby generating a broad weight distribution.
Still, similar to self-organized bistability, the system converges to a bistable state at the edge of the transition between high and low firing rates during development.
Here, this bistability remains even after the homeostatic feedback is switched off, because fluctuations from recurrent dynamics as well as external input are strong enough to induce transitions between both activity states.

The here identified mechanism of a stochastic state switching thereby presents a new perspective on so-called up-and-down states.
Up-and-down states are defined on the level of a single-neuron membrane potential that switches between states with higher membrane potential, resulting in spiking responses, and those with lower membrane potential~\cite{wilson_up_2008}.
While some of the aforementioned models utilize adaptation mechanisms to generate up-and-down states~\cite{millman_self-organized_2010,di_santo_self-organized_2016, buendia_self-organized_2020}, we here develop an alternative explanation:
If neurons homeostatically regulate their firing rates, a decreasing external input can result in emergent autocorrelations with potential functional benefits~\cite{zierenberg_homeostatic_2018, cramer_control_2020} until a point where bistability can emerge on the population level.
As a result of such (local) collective bistability, single neurons could switch between states of high and low synaptic input, which could in turn cause up-and-down states on the level of their membrane potentials.
Such up-and-down states have been observed mostly in striatal neurons~\cite{wilson_up_2008,stern_spontaneous_1997}, attributed to single-cell bistability, but also in cortical slices~\cite{cossart_attractor_2003} and neuronal cultures~\cite{vardi_simultaneous_2016}, where their origin is argued to be a collective (network) effect.
Our results provide new paths to test whether such observations of up-and-down states are connected to emergent bistability from homeostatic regulation via in-vitro assays or experiments with sensory deprivation.

Our general mean-field theory of fluctuation-induced bistability further presents a new paradigm for arbitrary finite systems with absorbing states and external drive.
Examples of such systems include collective dynamics in epidemic spread~\cite{pastor-satorras_epidemic_2015}, neural networks~\cite{beggs_neuronal_2003, chialvo_emergent_2010, wilting_25_2019}, ecosystems~\cite{scheffer_critical_2009, martin_eluding_2015}, and ultracold Rydberg atomic gases~\cite{helmrich_signatures_2020}, catalytic reactions on surfaces~\cite{ehsasi_steady_1989}, calcium dynamics in living cells~\cite{bar_discrete_2000}, or turbulence in liquid crystals~\cite{takeuchi_directed_2007, takeuchi_experimental_2009} and active nematics~\cite{doostmohammadi_onset_2017}.
Indeed, for some of these systems the phenomenon of bistability has been observed, e.g., as switching behavior in disease models~\cite{bottcher_critical_2017} or rate models of neural activity~\cite{van_meegen_large-deviation_2021}, for cellular automata with long-range interactions~\cite{pizzi_bistability_2021}, for CO oxidation~\cite{ertl_oscillatory_1991, suchorski_role_2018, wang_bistability_2019}, or as phase separation in active matter~\cite{martin_fluctuation-induced_2021, di_carlo_evidence_2022}.
With our work, we present fluctuation-induced bistability as a new paradigm that should be observable (or avoidable) if finite-size fluctuations are controlled accordingly.

%% file: content/acknowledgements.tex
\section*{Acknowledgments}

This work has received funding from the European Union Sixth Framework Programme ([FP6/2002-2006]) under grant agreement no 15879 (FACETS), the European Union Seventh Framework Programme ([FP7/2007-2013]) under grant agreement no 604102 (HBP), 269921 (BrainScaleS) and 243914 (Brain-i-Nets) and the Horizon 2020 Framework Programme ([H2020/2014-2020]) under grant agreement no 720270, 785907, and 945539 (HBP), the Deutsche Forschungsgemeinschaft (DFG, German Research Foundation) under Germany’s Excellence Strategy EXC 2181/1-390900948 (the Heidelberg STRUCTURES Excellence Cluster), the Helmholtz Association Initiative and Networking Fund [Advanced Computing Architectures (ACA)] under Project SO-092 as well as the Manfred St\"{a}rk Foundation.
MAM acknowledges the Spanish Ministry and Agencia Estatal de investigaci{\'o}n (AEI) through Project of I+D+i Ref. PID2020-113681GB-I00, financed by MICIN/AEI/10.13039/501100011033 and FEDER “A way to make Europe”, as well as the Consejer{\'\i}a de Conocimiento, Investigaci{\'o}n Universidad, Junta de Andaluc{\'\i}a and European Regional Development Fund, Project P20-00173 for financial support.
VP and JZ were supported by the Max Planck Society.
JZ received financial support from the Joachim Herz Stiftung and the Plan Propio de Investigaci\'on y Transferencia de la Universidad de Granada.
The authors acknowledge support by the state of Baden-W\"{u}rttemberg through bwHPC.

%% file: content/appendix.tex
\begin{figure}[t]
	\centering
	\input{content/routing.tex}
	\caption{%
		\textbf{The connectivity on BrainScaleS-2 is physically represented by two arrays of synapses.}
		The routing capabilities of BrainScaleS-2 are utilized such that both arrays can be treated as a larger virtual one.
		Input events $s_i^l$ enter this array together with recurrent events $t_i^k$ (blue and gray) from the left via synapse drivers (triangles).
		The latter forward the events to a whole row of synapses (circles).
		Each synapse locally filters incoming events and either transmits input events (red) or recurrent events (blue and gray) to its home neuron.
		Sparsity is implemented by silencing out synapses (black crosses).
		Homeostatic regulation is carried on by the on-chip \acrshort{ppu} by accessing neuronal firing rates $\nu_i$ to update synaptic weights in a row-wise parallel manner.
	}
	\label{fig:routing}
\end{figure}
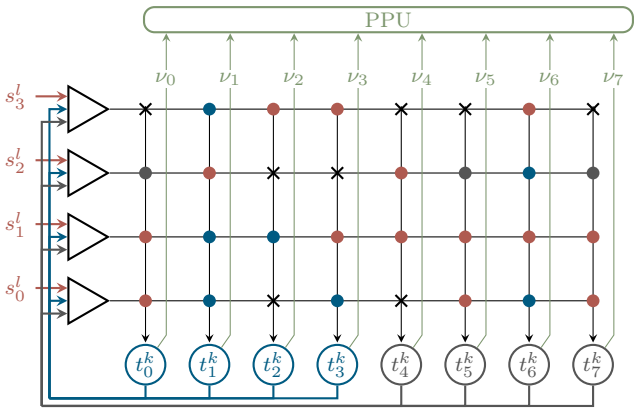

\section{Hardware details} \label{sec:appendix_hardware}

The connectivity on BrainScaleS-2 is physically represented by two arrays of synapses, each with a set of $256\times 256$ synapses.
Input spikes $s_i^l$ enter these arrays from the left via synapse drivers and are forwarded to a whole row of synapses.
Each synapse within this row locally filters incoming events, weights them according to a \SI{6}{\bit} weight and eventually transmits them to its home neuron.
The neurons are arranged in an additional row below the array of synapses.
Emitted neuronal spikes are injected back into the array via a flexible on-chip spike router.

Within this work, we exploit the routing capabilities of the BrainScaleS-2 system to unify both synapse arrays to a virtual array of size $256\times 512$ (\cref{fig:routing}).
The event filtering within each synapse located between synapse driver $i$ and neuron $j$ is used to either transmit the input events of spike source $i$ or the recurrent events of the neuron $i$ or $i + 256$, respectively.
We map our networks by configuring a random set of on average $K^\mathrm{rec}$ synapses per column of synapses to transmit recurrent events.
In addition, on average $K^\mathrm{in}$ randomly chosen synapses relay the input spike trains.
All remaining synapses are configured to transmit no events.

On BrainScaleS-2, the effect of each synapse, i.\,e.\, excitatory or inhibitory, is determined by the synapse drivers and is therefore a row-wise property within the synapse array.
For our networks, we program \SI{20}{\percent} of the synapse drivers (randomly selected) to be inhibitory.

The homeostatic plasticity is implemented on-chip by drawing on both \glspl{ppu} \citep{friedmann_demonstrating_2017}.
To that end, the number of emitted spikes is accessed and loaded into the \gls{simd} vector units of the \glspl{ppu} for subsequent weight update calculations.
Each vector unit allows to update a half-row of synaptic weights ($128$) synapses in parallel.
Calculations are performed with a precision of \SI{8}{\bit} in a fractional-saturation arithmetic.
The random numbers required to implement stochastic weight updates are directly drawn in parallel on-chip via dedicated accelerators.

\begin{figure*}[ht]
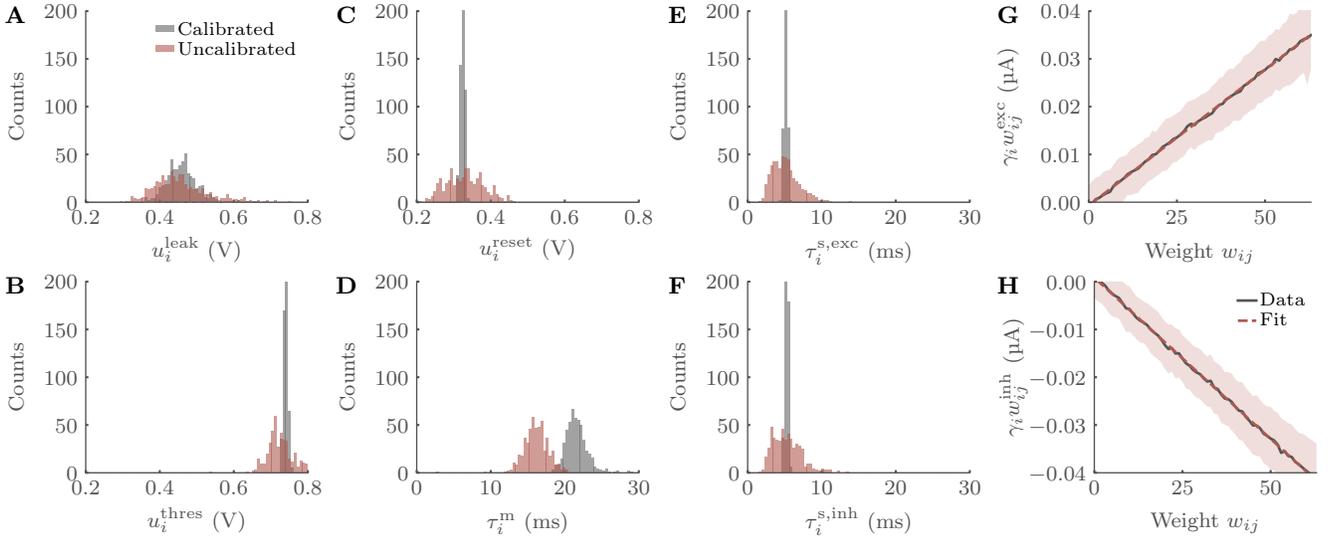

	\centering
	\begin{tikzpicture}
		\node[anchor=north west,inner sep=0pt] (a) at (0,0) {\input{build/figures/v_leak.pgf}};
		\node[anchor=north west,inner sep=0pt] (b) at (0,-3.6) {\input{build/figures/v_thres.pgf}};
		\node[anchor=north west,inner sep=0pt] (c) at (4.4,0) {\input{build/figures/v_reset.pgf}};
		\node[anchor=north west,inner sep=0pt] (d) at (4.4,-3.6) {\input{build/figures/tau_m.pgf}};
		\node[anchor=north west,inner sep=0pt] (e) at (8.8,0) {\input{build/figures/tau_syn_exc.pgf}};
		\node[anchor=north west,inner sep=0pt] (f) at (8.8,-3.6) {\input{build/figures/tau_syn_inh.pgf}};
		\node[anchor=north west,inner sep=0pt] (g) at (13.2,0) {\input{build/figures/epsp_height.pgf}};
		\node[anchor=north west,inner sep=0pt] (h) at (13.2,-3.6) {\input{build/figures/ipsp_height.pgf}};
		\node at ($(a.north west) + (0.2,-0.2)$) {\textbf{A}};
		\node at ($(b.north west) + (0.2,-0.2)$) {\textbf{B}};
		\node at ($(c.north west) + (0.2,-0.2)$) {\textbf{C}};
		\node at ($(d.north west) + (0.2,-0.2)$) {\textbf{D}};
		\node at ($(e.north west) + (0.2,-0.2)$) {\textbf{E}};
		\node at ($(f.north west) + (0.2,-0.2)$) {\textbf{F}};
		\node at ($(g.north west) + (0.2,-0.2)$) {\textbf{G}};
		\node at ($(h.north west) + (0.2,-0.2)$) {\textbf{H}};
	\end{tikzpicture}
	\caption{%
		\textbf{Parameter distributions on the BrainScaleS-2 system.}
		Calibration routines allow to reduce the parameter spread between circuit instances by drawing on the configurability of BrainScaleS-2.
		The calibration targets for \textbf{(A)} the leak potential $u^\mathrm{leak}_i$ and \textbf{(B)} the threshold potential $u^\mathrm{thres}_i$ are chosen such that their distance is as high as possible.
		\textbf{(C)} The target for the reset potential is chosen slightly below $u^\mathrm{leak}_i$.
		\textbf{(D)} The excitatory synaptic time constants $\tau^\mathrm{s,exc}_i$ and \textbf{(E)} the inhibitory synaptic time constants $\tau^\mathrm{s,inh}_i$ are calibrated to coincide.
		\textbf{(F)} The membrane time constant $\tau^\mathrm{m}_i$ exceeds the synaptic time scales.
		Linear fits to measurements of the \acrshort{psp} height for various weight values $w_{ij}$ allow to estimate \textbf{(G)} the excitatory weight scaling factors $\gamma_i^\mathrm{exc}$ as well as \textbf{(H)} the inhibitory weight scaling factor $\gamma_i^\mathrm{inh}$.
	}
	\label{fig:parametrization}
\end{figure*}

\section{Calibration and parametrization \label{sec:parametrization}}

The fabrication induced device variations of the analog neuromorphic substrate are mitigated by calibration routines.
Here, we utilize bisection methods to adjust the neuro-synaptic parameters inferred from recorded traces to desired targets (\cref{fig:parametrization}).
Afterwards, the resulting \gls{lif} neuron parameters are measured and their mean as well as standard deviations are used for the parametrization of the equivalent software models.
To align the impact of a single spike on all downstream neurons on hard- and in software, we characterize the \gls{psp} height as a function of the configured weight value $w_{ij}$.
In more detail, we obtained the weight scaling factor $\gamma_i$ in \cref{eq:synapse} by fitting the ideal solution of \cref{eq:membrane}
\begin{align}
	u_i(t) = & \, u^\mathrm{leak} + \frac{\tau^\mathrm{m}\cdot\tau^\mathrm{s}\cdot \gamma_i w_{ij}}{C^\mathrm{m} (\tau^\mathrm{s} - \tau^\mathrm{m})} \Theta\left(t - t_j^0\right) \nonumber\\
	&\cdot\left[\exp{\left(-\frac{t-t_j^0}{\tau^\mathrm{s}}\right)} - \exp{\left(-\frac{t-t_j^0}{\tau^\mathrm{m}}\right)}\right] \, ,
	\label{eq:lif_solution}
\end{align}
to recordings of each neuron's membrane potential $u_i(t)$ in response to a single stimulating event $t_j^0$ relayed over a single synapse with weight $w_{ij}$.
To ensure stable fits, we fix all fit parameters to the calibration target values except for the leak potential $u^\mathrm{leak}$ and our estimate of $\gamma_i w_{ij}$ that we call $y$.
From the linear fit $y = \gamma_i w_{ij}$, we then obtain an estimate of $\gamma_i$ for excitatory and inhibitory weights respectively (\cref{fig:parametrization} G and H).
All estimated parameters are summarized in \cref{tab:parameters}.

\begin{table}[b]
	\centering
	\caption{%
		\textbf{Model parameters.}
		All parameters are given in the equivalent biological time domain.
		The errors indicate the standard deviation.
	}
	\label{tab:parameters}
	\input{tables/parameter}
\end{table}

\begin{figure}[ht]
	\centering
	\begin{tikzpicture}
		\node[anchor=north west,inner sep=0pt] (a) at (0,0){\import{build/figures/}{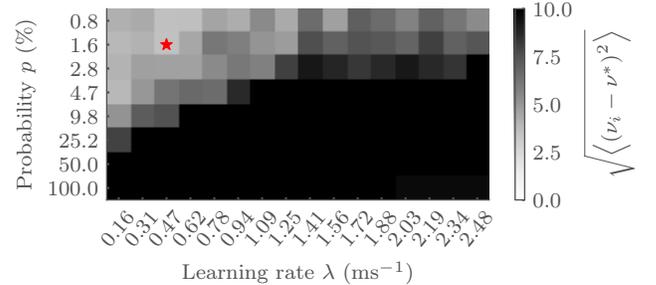}};
	\end{tikzpicture}
	\caption{%
		\textbf{Parametrization of the homeostatic regulation.}
		The homeostatic regulation is parametrized by the update acceptance probability $p$ as well as the learning rate $\lambda$.
		Shown is the variance of the firing rate of each neuron $\nu_i$ with respect to the target rate $\nu^\ast$, $\sqrt{\langle \left(\nu_i - \nu^\ast\right)^2\rangle}$, averaged over \num{100} experiments for an input rate of $h=\SI{0.6}{\kilo\hertz}$.
		The configuration used within the experiments presented in the main part is highlighted by the red star.
	}
	\label{fig:stability}
\end{figure}
\begin{figure*}[ht]
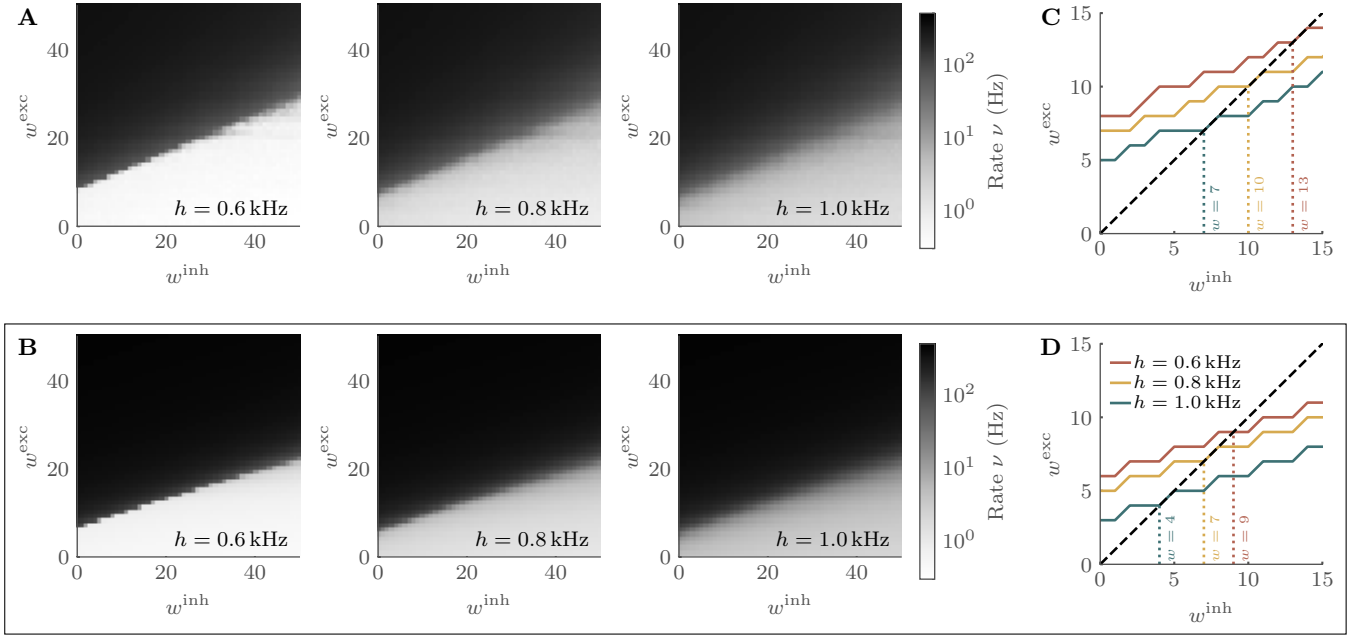

	\centering
	\begin{tikzpicture}
		\node[anchor=north west,inner sep=0pt] (a) at (0,0){\import{build/figures/}{phase_full_emul_60.pgf}};
		\node[anchor=north west,inner sep=0pt] (c) at (4.0,0) {\import{build/figures/}{phase_full_emul_80.pgf}};
		\node[anchor=north west,inner sep=0pt] (e) at (8.0,0.0) {\import{build/figures/}{phase_full_emul_100.pgf}};
		\node[anchor=north west,inner sep=0pt] (x) at (12.0,-0.1) {\import{build/figures/}{phase_full_emul_colorbar.pgf}};
		\node[anchor=north west,inner sep=0pt] (g) at (13.6,0.0) {\import{build/figures/}{phase_full_emul_cross.pgf}};

		\node[anchor=north west,inner sep=0pt] (b) at (0.0,-4.4) {\import{build/figures/}{phase_full_sim_60.pgf}};
		\node[anchor=north west,inner sep=0pt] (d) at (4.0,-4.4) {\import{build/figures/}{phase_full_sim_80.pgf}};
		\node[anchor=north west,inner sep=0pt] (f) at (8.0,-4.4) {\import{build/figures/}{phase_full_sim_100.pgf}};
		\node[anchor=north west,inner sep=0pt] (y) at (12.0,-4.5) {\import{build/figures/}{phase_full_sim_colorbar.pgf}};
		\node[anchor=north west,inner sep=0pt] (h) at (13.6,-4.4) {\import{build/figures/}{phase_full_sim_cross.pgf}};

		\node at ($(a.north west) + (0.2,-0.2)$) {\textbf{A}};
		\node at ($(b.north west) + (0.2,-0.2)$) {\textbf{B}};
		\node at ($(g.north west) + (0.2,-0.2)$) {\textbf{C}};
		\node at ($(h.north west) + (0.2,-0.2)$) {\textbf{D}};
		\node[draw,fit=(b) (h)] (box_simulation) {};
	\end{tikzpicture}
	\caption{%
		\textbf{Phase diagrams of networks with homogeneous and static weights.}
		\textbf{(A)} The firing rates $\nu$ for three exemplary input rates $h$ are comparable between hardware \textbf{(A)} and software \textbf{(B)} implementations.
		However, there is a small shift of the transitions from high firing rates $\nu$ to intermediate firing rates between both.
		For each value of the configured inhibitory weight $w^\mathrm{inh}$, the firing rate is closest to \SI{10}{\hertz} for a coinciding excitatory weight $w^\mathrm{exc}_\mathrm{cross}$ for both, emulation \textbf{(C)} as well as simulation \textbf{(D)}.
	}
	\label{fig:phase_full}
\end{figure*}

\section{Parametrization of the homeostatic regulation \label{sec:hom_parametrization}}

The homeostatic regulation as given by \cref{eq:rule} comes with two independent parameters: the learning rate $\lambda$ as well as the update acceptance probability $p$.
We obtained optimal parameters by performing a grid search for $h = \SI{0.6}{\kilo\hertz}$ and assessing the variance of the firing rate of all neurons $\nu_i$ with respect to the target rate $\nu^\ast$, i.\,e\,. $\sqrt{\langle \left(\nu_i - \nu^\ast\right)^2\rangle}$.
For a broad range of parameters, most of the \gls{lif} neurons emit spikes at a rate resembling the target rate (\cref{fig:stability}).
Only for low values of $\lambda$ and high values of $p$, the firing rate systematically deviates due to the integer arithmetic used for weight update calculations on the neuromorphic system.

Most notably, we also used the determined optimal parameters within our software simulations.
This pursued strategy renders extensive parameter sweeps in software superfluous and moreover showcases the benefits of the accelerated analog emulation of neuro-synaptic dynamics due to the referenced efficiency in terms of speed and power consumption.

\begin{figure*}[ht]
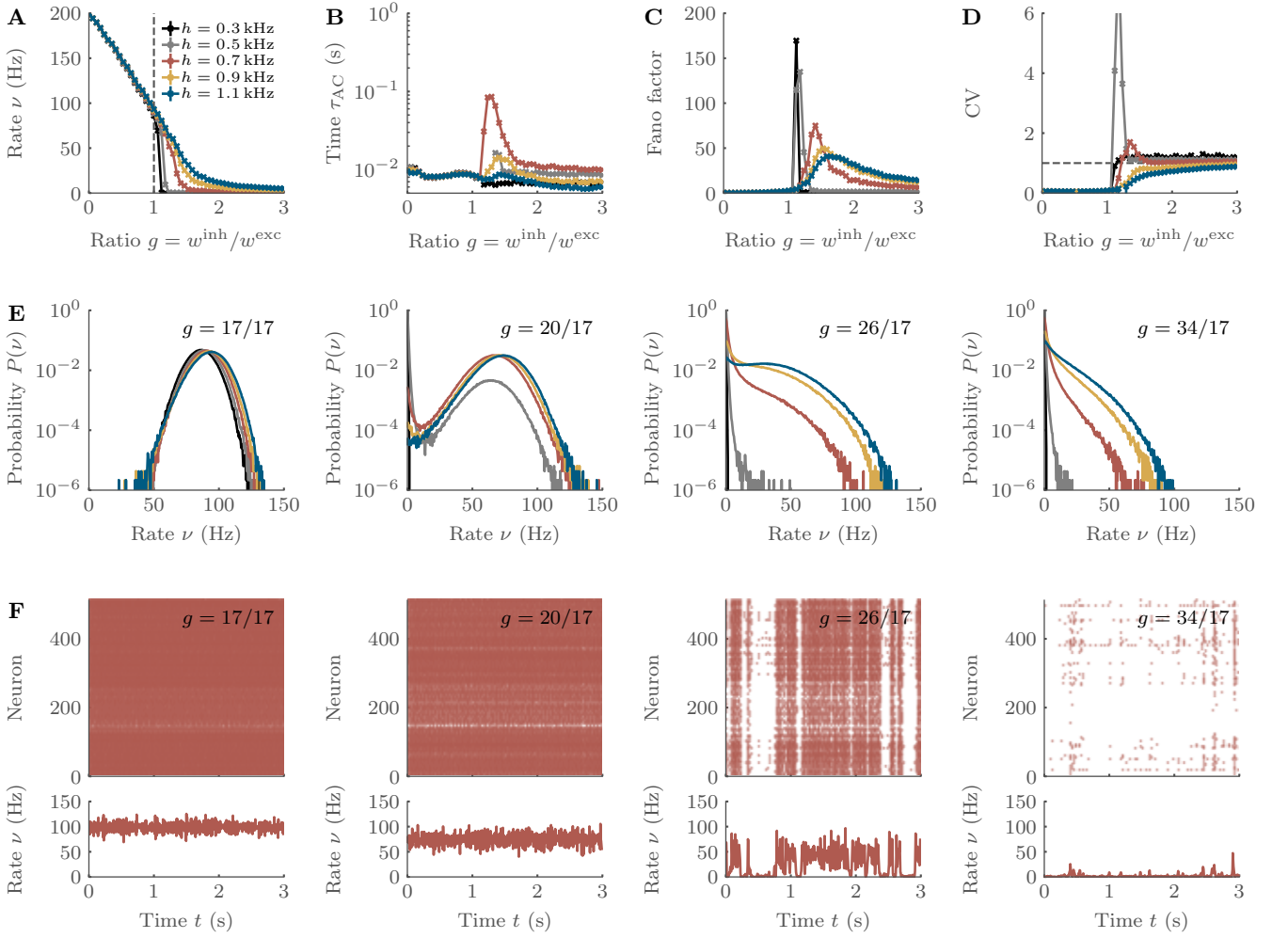

	\centering
	\begin{tikzpicture}
		\node[anchor=north west,inner sep=0pt] (a) at (0,0){\import{build/figures/}{phase_rate.pgf}};
		\node[anchor=north west,inner sep=0pt] (b) at (4.5,0) {\import{build/figures/}{phase_taus.pgf}};
		\node[anchor=north west,inner sep=0pt] (c) at (9.0,0.0) {\import{build/figures/}{phase_fano.pgf}};
		\node[anchor=north west,inner sep=0pt] (d) at (13.5,0.0) {\import{build/figures/}{phase_cv.pgf}};

		\node[anchor=north west,inner sep=0pt] (e) at (0.0,-4.2) {\import{build/figures/}{phase_activity_distribution_0.pgf}};
		\node[anchor=north west,inner sep=0pt] (f) at (4.5,-4.2) {\import{build/figures/}{phase_activity_distribution_1.pgf}};
		\node[anchor=north west,inner sep=0pt] (g) at (9.0,-4.2) {\import{build/figures/}{phase_activity_distribution_2.pgf}};
		\node[anchor=north west,inner sep=0pt] (h) at (13.5,-4.2) {\import{build/figures/}{phase_activity_distribution_3.pgf}};

		\node[anchor=north west,inner sep=0pt] (i) at (0,-8.4){\import{build/figures/}{phase_raster_0.pgf}};
		\node[anchor=north west,inner sep=0pt] (j) at (4.5,-8.4){\import{build/figures/}{phase_raster_1.pgf}};
		\node[anchor=north west,inner sep=0pt] (k) at (9.0,-8.4){\import{build/figures/}{phase_raster_2.pgf}};
		\node[anchor=north west,inner sep=0pt] (l) at (13.5,-8.4){\import{build/figures/}{phase_raster_3.pgf}};
		\node at ($(a.north west) + (0.2,-0.2)$) {\textbf{A}};
		\node at ($(b.north west) + (0.2,-0.2)$) {\textbf{B}};
		\node at ($(c.north west) + (0.2,-0.2)$) {\textbf{C}};
		\node at ($(d.north west) + (0.2,-0.2)$) {\textbf{D}};
		\node at ($(e.north west) + (0.2,-0.2)$) {\textbf{E}};
		\node at ($(i.north west) + (0.2,-0.2)$) {\textbf{F}};
	\end{tikzpicture}
	\caption{%
		\textbf{Dynamics of networks with homogeneous and static weights.}
		\textbf{(A)} The transitions from high firing rates $\nu$ to intermediate firing rates occurs in the vicinity of $g\approx 1$ for decreasing $h$.
		\textbf{(B)} The autocorrelation time $\tau_\mathrm{AC}$ and \textbf{(C)} the Fano factor are estimated based on the population activity obtained with a binsize of \SI{5}{\milli\second} and show peaks around the finite-size transition.
		\textbf{(D)} Average Coefficient of variation of single neuron inter-spike intervals suggests regular spiking for small $g$ and irregular spiking for large $g$.
		\textbf{(E)} Distributions of population rates shown in (A) for slices of different $g$ show the transition from regular firing at high rates to irregular firing at low rates.
		\textbf{(F)} Example snapshots of spike rater plots and population rate for $h=\SI{0.7}{\kilo\hertz}$ for different $g$.
	}
	\label{fig:phase}
\end{figure*}

\section{Phase diagrams of networks with homogeneous and static weights \label{sec:phase_diagrams}}

To understand why we can observe fluctuating or bistable dynamics in networks with homeostatically regulated weights despite apparent excitation dominance (cf.\ \cref{fig:chip} and \cref{fig:time_scale}), we study here the phase diagram of comparable networks with homogeneous and static weights.
Due to small fluctuations in the transition point for different realizations of small networks, we focus on a single network realization and split the measurement into \num{200} blocks of length $\SI{30}{\second}$.
To ensure spiking activity even for low input strengths $h$, we initially increased $h$ for \SI{5}{\second} and subsequently let the networks equilibrate for another \SI{5}{\second}.

We first perform a full sweep over the $w_\mathrm{exc}-w_\mathrm{inh}$ plane on both the BrainScaleS-2 and a corresponding software simulations for three exemplary input strengths (\cref{fig:phase_full} A and B).
While the overall trend of the firing rates in emulation and simulation is quite comparable, the transition from low to high firing rates is clearly shifted.
We attribute these remaining differences to (i) the fact that our simulations do not capture correlations in the variability of parameters, but instead implement uncorrelated Gaussian noise, and (ii) to additional saturation effects within the analog circuits of BrainScaleS-2.

When we consider as a proxy for the transition between high-firing and low-firing phase the line where $\nu\approx 10$, we notice that this transition occurs for $w^\mathrm{exc} \approx \left(w^\mathrm{inh} + o\right)/s$ with $o>0$ an $h$-dependent offset (\cref{fig:phase_full} C and D).
Hence, this transition does not occur for a fixed inhibition-excitation ratio, $g=w^\mathrm{inh}/w^\mathrm{exc} \approx s - o/w^\mathrm{exc}$.
Instead, $g$ depends non-trivially on the excitatory coupling as well as on the parameters $s$ and $o$, which further depend on the input rate $h$ and the specific choice of input coupling (see \cref{sec:methods}).

When we now interpret our symmetric homeostatic rule to only allow identical couplings $w_\mathrm{exc}=w_\mathrm{inh}$ (\cref{fig:phase_full} C and D, black dashed line), then homeostatic plasticity should adjust the weights to the intersection between the transition lines and the unit line.
While this is strongly simplified, it approximately recovers the range of resulting mean weights that we find for the homeostatically-regulated neuromorphic chip (\cref{fig:chip}) and simulations (\cref{fig:simulation_comparison_homeostasis}).

To characterize the dynamical phases of high- and low-firing rates, we focus the special cutplane of $w^\mathrm{exc}=w^\mathrm{in}=17$ on the BrainScaleS-2 system.
We record the mean neuron firing rate, the integrated autocorrelation time, the network Fano factor, as well as the coefficient of variation (CV) of inter-spike intervals as a function of the inhibition dominance, $g=w^\mathrm{inh}/w^\mathrm{exc}$.
The integrated autocorrelation time is estimated from integrating the autocorrelation function $C(t^\prime)$, cf.\, \cref{eq:correlation}.
We follow the common convention~\cite{grotendorst_statistical_2002} to define $\tau_\mathrm{int}=\Delta t[\frac{1}{2}+\sum_{l=1}^{l_\mathrm{max}} C(l)]$, where $l_\mathrm{max}$ is self-consistently obtained as the first $l$ for which $l > 6\,\tau_\mathrm{int}(l)$.
This reliably estimates the scale of temporal correlations for fully sampled systems and did not become unstable due to the typical oscillations in the autocorrelation function observed for networks of \gls{lif} neurons.
Since the communication bottleneck of the hardware constrained long samples for high firing rates, we partitioned each recording into $L=\num{200}$ chunks of size $T=\SI{30}{\second}$ and estimated for each chunk the moments as averages, i.e., $\overline{\nu(t)}_l=\frac{1}{T}\sum_t \nu(t)$ and $\overline{\nu(t)\nu(t+t^\prime)}_l=\frac{1}{T}\sum_t \nu(t)\nu(t+t^\prime)$.
To avoid finite-data biases~\cite{marriott_bias_1954}, we then first obtained the best estimates of the mean $\overline{\nu(t)}=\frac{1}{L}\sum_l\overline{\nu(t)}_l$ and analogously of the correlation term, to then estimate the covariance as $\text{Cov}[\nu(t+t^\prime)\nu(t)]=\overline{\nu(t)\nu(t+t^\prime)} - \overline{\nu(t)}^2$.
Similarly, we estimate the network Fano factor on the population rate as the ratio between variance and mean, i.e., $F=(\overline{\nu^2(t)}-\overline{\nu(t)}^2)/\overline{\nu(t)}$, and the CV as average across neurons, i.e. $\text{CV}=\frac{1}{N}\sum_i \sqrt{\overline{\delta t^2}_i-\overline{\delta t}^2_i}/\overline{\delta t}_i$ with inter-spike-intervals $\delta t_i^j$ of neuron $i$.

We find that for the considered setup with an \gls{ei} input layer, the transition from high firing rates to low firing rates is reminiscent of a regular-to-irregular transition (\cref{fig:phase}A).
For the special choice $w^\mathrm{exc}=w^\mathrm{in}$, the transition occurs at $g\approx 1$ for $h\to0$, where the dynamic phase in the inhibition-dominated regime appears absorbing despite non-vanishing input due to the small system size.
In the vicinity of the $h$-dependent transition, we observe peaks in the autocorrelation time (\cref{fig:phase}B), which we expect to vanish due to the absorbing state in the limit of $h\to 0$ and $N\to\infty$~\cite{zierenberg_notitle_nodate}.
We find that the network Fano factor, estimated from the population activity with $\Delta t = \SI{5}{\milli\second}$ (\cref{fig:phase}C), is zero in the regular phase and low in the irregular phase, separated again by a peak that shifts towards $g=1$ with decreasing $h$ and becomes narrower.
Last, we observe the average coefficient of variation of single-neuron inter-spike intervals to change from $\text{CV}\approx 0$ for $g<1$, indicating regular spiking, to $\text{CV}\approx 1$ above the transition, indicating irregular spiking (\cref{fig:phase}D).

To illustrate the dynamic phases of regular and irregular activity, we show distributions of population rates (\cref{fig:phase}E) as well as spike raster plots and the time evolution of the population rate (\cref{fig:phase}F).
For $g=17/17=1$ we find all $h$ in a stable active state.
For $g=20/17\approx 1.2$ $h=\SI{0.3}{kHz}$ is already in the quiescent state, while $h=\SI{0.5}{kHz}$ shows strong variance between high rate and low rates that hinder estimation of autocorrelation times, but all other $h$ remain mostly in the stable active state.
For $g=26/17\approx 1.5$, we observe highest autocorrelations for $h=\SI{0.7}{kHz}$ due to strong fluctuation-driven excursions into the high-firing rate regime.
Further increasing $g$ also causes the other $h$ to fall into low-firing-rate states, where for small $h$ the state appears absorbing with practically no population activity following upon the few external perturbations.
Note that single points of the phase diagrams cannot be directly compared to the results after homeostatic regulation, which results in heterogeneous weight distributions (cf. \cref{fig:chip}E), because we here fix $w^\mathrm{exc}=w^\mathrm{in}$.

\begin{figure}[ht]
	\centering
	\begin{tikzpicture}
		\node[anchor=north west,inner sep=0pt] at (0,0){\import{build/figures/}{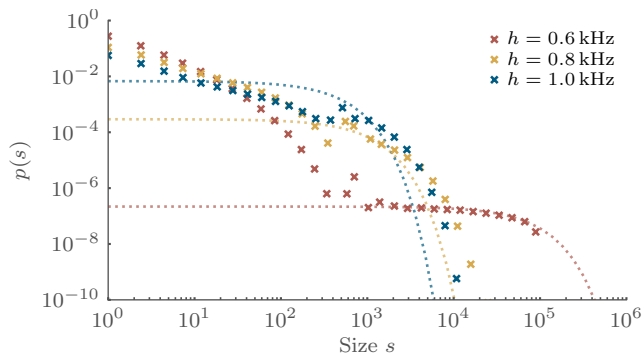}};
	\end{tikzpicture}
	\caption{%
        \textbf{Tail of avalanche-size distribution from homeostatically regulated neuromorphic networks can be explained by the timescales of a \gls{hmm}.}
        Empirical avalanche-size distributions for different $h$ in a log-binned representation show now power-law shape.
        In contrast, the cutoff scale of their tails are well described by estimates based on a \gls{hmm} with data-constrained parameters.
	}
	\label{fig:avalanches}
\end{figure}

\section{Avalanche analysis} \label{sec:avalanches}
To verify that the autocorrelations we observe are not a result of critical fluctuations, we investigate the distribution of avalanche sizes.
For (self-organized) critical systems, one would expect avalanche sizes $s$ to be scale free~\cite{bak_self-organized_1988, pruessner_self-organised_2012}, i.e., an avalanche-size distribution $p(s)$ that can be described by a power law.

Here, we follow the convention to estimate avalanches from a time-discrete firing rate $\nu(t)$~\cite{beggs_neuronal_2003, zeraati_self-organization_2021}.
To constrain the temporal bin size to causal activity propagation, we estimate the spike delay from the solution of the \gls{lif} equation, \cref{eq:lif_solution}.
More specifically, the peak of the \gls{epsp} is an estimate of the maximal time until a certain spike can causally induce a threshold crossing.
We obtain the peak time of the \gls{epsp} from the condition $du / dt = 0$ in \cref{eq:lif_solution}, which together with the spike delay yields
\begin{equation}
	\tau^\mathrm{tot} = \tau^\mathrm{d} + \frac{\tau^\mathrm{m}\tau^\mathrm{s} \log{\left(\frac{\tau^\mathrm{m}}{\tau^\mathrm{s}}\right)}}{\tau^\mathrm{m} - \tau^\mathrm{s}} \, .
\end{equation}
Since $\tau^\mathrm{tot}$ sets an upper estimate of the causal delay, we here set the bin size for avalanche detection to $\Delta t = \tau^\mathrm{tot} / 2 = \SI{5}{\milli\second}$, in agreement with our previous time discretization.
An avalanche is then defined as the number of spikes in consecutive non-empty bins in $\nu(t)$, for which we measured $L=\num{200}$ chunks of size $T=\SI{100}{\second}$.

To verify that the resulting avalanche-size distributions are not a result from critical fluctuations, we compare with predicted distributions based on our \gls{hmm} (\cref{fig:avalanches}).
The \gls{hmm} approximates $\nu(t)$ with a Markov jump process that switches between periods of high and low firing rates.
For this, the process assumes state durations to be exponentially distributed (which we confirmed for low $h$, not shown).
Introducing a life-time $T_+$ and a conditional average rate $\nu_{+}$ for the high-firing state, we can approximate large avalanches sizes as $s=\nu_{+}T_+$ with $p(s) \propto p(T_+=s/\nu_+)$.
Since we do not have an unbiased estimate of the fraction of avalanches in the low-firing rate, we constrain the amplitude from a fit to the tail of the empirical avalanche-size distribution (\cref{fig:avalanches}, dashed lines).
The predicted exponential avalanche-size distributions capture the cutoff scale of large avalanche sizes with increasing match for lower $h$.
We thus conclude that the empirical avalanche-size distributions show no sign of scale-free avalanches but are captured by the dynamics of the \gls{hmm}.

\section{Simulation of mean-field model} \label{sec:appendix_meanfield}

To simulate the time evolution of the mean-field model, one needs special care to avoid negative densities from numerical imprecisions that would render the multiplicative noise imaginary~\cite{dornic_integration_2005}.
In short, the steps involve first evaluating an exact solution of the noise and linear terms and then an Euler integration of the remaining quadratic term.
The precise mean-field equation we solve is
\begin{align}
    \dot{\rho}(t) = &h - \left(\tau_\mathrm{MF}-\alpha +h[1+\beta]\right)\rho(t)+ \sigma\sqrt{\rho(t)/N}\eta(t)\nonumber\\
    &-b\rho^2(t) \, .
\end{align}
This equation is decomposed into the linear term plus noise (first line), for which one can obtain an analytical solution, and the higher-order term (second line), which can be trivially integrated.

For the square-root noise plus linear term, i.e., $\dot{\rho}(t)=h+a\rho +\tilde{\sigma}\sqrt{\rho}\eta$, starting from $\rho(t)=\rho_0$ one knows the solution of the Fokker Planck equation for time $t+\Delta t$ is~\cite{feller_two_1951}
\begin{align}
    P(\rho,t+\Delta t) = \lambda e^{\lambda\left(\rho_0\omega+\rho\right)}\left[\frac{\rho}{\rho_0\omega}\right]^{\mu/2} I_\mu\left(2\lambda\sqrt{\rho_0\rho\omega}\right),
\end{align}
where $I_\mu$ is the modified Bessel function of order $\mu$, $\omega=e^{a\Delta t}$, $\lambda=2a/\sigma^2\left(\omega-1\right)$, and $\mu=-1+2h/\sigma^2$.
Using a Taylor-series expansion of the Bessel function, it was shown in Ref.~\cite{dornic_integration_2005} that rewriting $P(\rho,t+\Delta t)$ implies the density after $\Delta t$ can be simply drawn from the mixture
\begin{align}\label{eq:step1}
    \rho^\ast = \text{Gamma}\left[\mu + 1 + \text{Poisson}\left[\lambda\rho_0\omega\right]\right]/\lambda.
\end{align}

We thus evolve $\rho(t)$ in discrete time steps $\Delta t$ in two steps~\cite{dornic_integration_2005}:
First, we generate from $\rho_0=\rho(t)$ the stochastic solution $\rho^\ast$ using \cref{eq:step1}.
Second, we integrate the remaining term as $\rho(t+\Delta t)=\rho^\ast/(1+\rho^\ast b \Delta t)$.

For the example we show in \cref{fig:theory}, we used $\Delta t=10^{-7}$, $\tau=10$, $\alpha=19$, $\beta=10$, $b=12$, $\sigma=40$, and $N=512$. For further details we refer to the available code~\cite{noauthor_benjamincramerneuromorphic-bistability_nodate}.

%% file: content/routing.tex
\definecolor{blue}{HTML}{1f77b4}%
\definecolor{red}{HTML}{d62728}%
\definecolor{green}{HTML}{2ca02c}%
\definecolor{yellow}{HTML}{fee23e}%
\definecolor{hidden}{HTML}{005b82}%
\definecolor{input}{HTML}{af5a50}%
\definecolor{ppu}{HTML}{7d966e}%
\definecolor{output}{HTML}{555555}%
\tikzset{silent/.style={cross out, draw, 
         minimum size=2*(3pt-\pgflinewidth), 
	 inner sep=0pt, outer sep=0pt, thick}}
\tikzset{input_synapse/.style={circle,minimum size=0.17cm,inner sep=0pt,fill=input}}%
\tikzset{recurrent0_synapse/.style={circle,minimum size=0.17cm,inner sep=0pt,fill=hidden}}%
\tikzset{recurrent1_synapse/.style={circle,minimum size=0.17cm,inner sep=0pt,fill=output}}%
\pgfmathdeclarerandomlist{MyRandomSynapses}{%
    {input_synapse}%
    {recurrent0_synapse}%
    {recurrent1_synapse}%
    {silent}%
}%
\tikzset{block/.style={font={\rmfamily\footnotesize},align=center}}%
\tikzset{box/.style={draw=black!90}}%
\tikzset{block label/.style={fill=white,font={\rmfamily\footnotesize},inner sep=0.05cm}}%
\tikzset{%
	neuron/.style = {%
		draw=black,%
		circle,%
		inner sep=0pt,%
		minimum width=0.3cm%
	},%
	driver/.style = {%
		minimum height=0.7cm,%
		draw=black,%
		regular polygon,%
		regular polygon sides=3,%
		shape border rotate=-90,%
		inner sep=0pt%
	},%
}%
\begin{tikzpicture}[
		x=1.7cm,
		y=1.7cm,
	    	anchor=center,
        ]
        \pgfdeclarelayer{background layer}
        \pgfsetlayers{background layer,main}

	\begin{scope}
		\foreach \x in {0,1,...,7} {
			\pgfmathparse{100*(\x>3)}
			\colorlet{currentcolor}{output!\pgfmathresult!hidden}
			
			\node[neuron,currentcolor,thick,inner sep=1pt] (nrn \x) at (0.8 + \x*0.5,0.35) {\fontsize{8}{8}\selectfont $t_\x^k$};
			\draw[stealth-] (nrn \x.north) ++ (0.0,0.01) -- ++(0.0,1.85);
		}

		\foreach \y [evaluate=\y as \z using {int(\y + 4)}] in {0,1,...,3} {
			\node[driver,thick] (drv \y) at ($(nrn 0) + (-0.5,0.5 + \y*0.5)$) {};
			\draw (drv \y.east) -- ++(3.8,0.0);
			\draw[stealth-,input,thick] ($(drv \y.center) + (-0.11,0.10)$) -- ++(-0.25,0.0) node[anchor=east] {\fontsize{8}{8}\selectfont $s_\y^l$};

			\foreach \x in {0,1,...,7} {
				\pgfmathrandomitem{\RandomSynapse}{MyRandomSynapses}
				\draw (drv \y -| nrn \x) node[\RandomSynapse] {};
			}
		}

		\foreach \x in {0,1,2,3} {
			\draw[hidden,thick] (nrn \x.south) -- ++(0.0,-0.1) coordinate (tmp) -- (tmp -| drv 0.west) -- ++(-0.14,0.0) coordinate (sammelpunkt);
		}
		
		\foreach \y in {0,1,...,3} {
			\draw[-stealth,hidden,thick] (sammelpunkt) -- ($(sammelpunkt |- drv \y) + (0.0,0.00)$) coordinate (tmp) -- (tmp -| drv \y.west);
		}
		
		\foreach \x in {4,5,6,7} {
			\draw[output,thick] (nrn \x.south) -- ++(0.0,-0.16) coordinate (tmp) -- (tmp -| drv 0.west) -- ++(-0.2,0.0) coordinate (sammelpunkt);
		}
		
		\foreach \y in {0,1,...,3} {
			\draw[-stealth,output,thick] (sammelpunkt) coordinate (tmp) -- ($(tmp |- drv \y) - (0.0,0.1)$) coordinate (tmp) -- (tmp -| drv \y.west);
		}

		\node[rectangle,thick,draw,ppu,rounded corners,inner sep=2pt,minimum width=6.5cm] (ppu) at ($(nrn 0) + (1.9,2.7)$) {\fontsize{8}{8}\selectfont PPU};
		\foreach \x in {0,1,...,7} {
			\draw[ppu,-stealth] (nrn \x.55) -- ++(55:0.12) coordinate (tmp) -- (tmp |- ppu.south) node[pos=0.85, fill=white, inner sep=1] {\fontsize{8}{8}\selectfont $\nu_\x$};
		}
	\end{scope}
\end{tikzpicture}

%% file: tables/parameter.tex
\footnotesize
\begin{ruledtabular}
	\begin{tabular}{llr}
		\textbf{Parameter} 			& \textbf{Symbol}		& \textbf{Value} \\
		\hline
		Membrane capacitance 			& $C^\mathrm{m}$		& \SI{2.4(2)}{\pico\farad} \\
		Threshold potential			& $u^\mathrm{thresh}$		& \SI{741(06)}{\milli\volt} \\
		Leak potential				& $u^\mathrm{leak}$		& \SI{458(43)}{\milli\volt} \\
		Reset potential				& $u^\mathrm{reset}$		& \SI{324(06)}{\milli\volt} \\
		Membrane time constant			& $\tau^\mathrm{m}$		& \SI{21.5(15)}{\milli\second} \\
		Excitatory synaptic time constant	& $\tau^\mathrm{s,exc}$		& \SI{5.3(3)}{\milli\second} \\
		Inhibitory synaptic time constant	& $\tau^\mathrm{s,inh}$		& \SI{5.4(2)}{\milli\second} \\
		Synaptic delay				& $\tau^\mathrm{d}$		& \SI{1.0(1)}{\milli\second} \\
		Refractory period			& $\tau^\mathrm{ref}$		& \SI{2.0}{\milli\second} \\
		Exciatotry weight scaling factor	& $\gamma^\mathrm{exc}$		& \SI{0.57(10)}{\nano\ampere} \\
		Inhibitory weight scaling factor	& $\gamma^\mathrm{inh}$		& \SI{0.67(10)}{\nano\ampere} \\
		\hline
		Recurrent synapses per neuron		& $K^\mathrm{rec}$		& \SI{100}{} \\
		Neurons					& $N$				& \SI{512}{} \\
		Input weight				& $w^\mathrm{in}$		& \SI{17}{} \\
		\hline
		Learning rate				& $\lambda$			& \SI{0.46875}{} \\
		Target rate				& $\nu^\ast$			& \SI{10}{\hertz} \\
		Update probability			& $p$				& \SI{2.3}{\percent} \\
		Number of updates			& $n$				& \SI{1000}{} \\
		\hline
		External rate				& $\nu^\mathrm{ext}$		& \SI{10}{\hertz} \\
		\hline
		Static experiment duration		& $T$				& \SI{100}{\second} \\
	\end{tabular}
\end{ruledtabular}

%% file: content/si.tex
\begin{figure*}[ht]
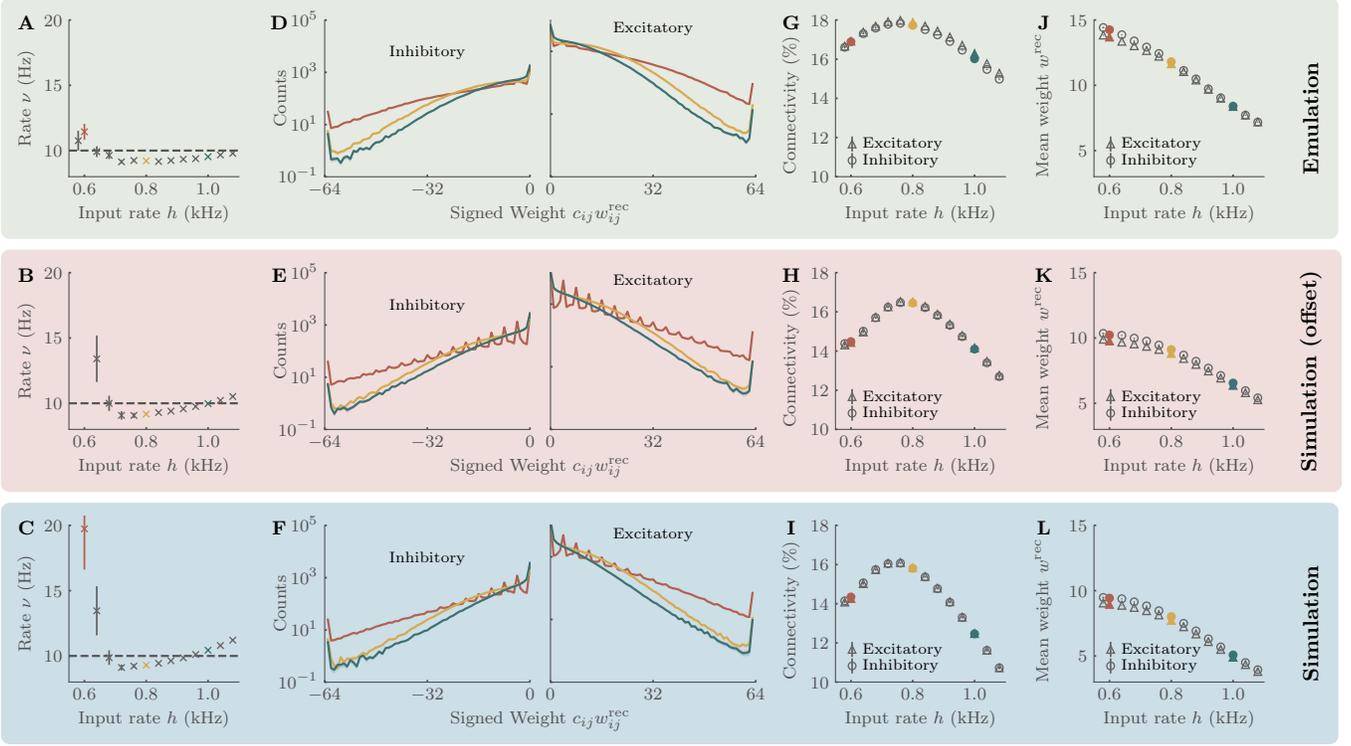

	\centering
	\scalebox{0.82}{
	\begin{tikzpicture}
		\node[anchor=north west,inner sep=0pt] (a) at (0,0) {\input{build/figures/si_emul_final_rate.pgf}};
		\node[anchor=north west,inner sep=0pt] (b) at (0,-4.1) {\input{build/figures/si_sim_offset_final_rate.pgf}};
		\node[anchor=north west,inner sep=0pt] (c) at (0,-8.2) {\input{build/figures/si_sim_final_rate.pgf}};

		\node[anchor=north west,inner sep=0pt] (d) at (4.1,0) {\input{build/figures/si_emul_weight_histogram.pgf}};
		\node[anchor=north west,inner sep=0pt] (e) at (4.1,-4.1) {\input{build/figures/si_sim_offset_weight_histogram.pgf}};
		\node[anchor=north west,inner sep=0pt] (f) at (4.1,-8.2) {\input{build/figures/si_sim_weight_histogram.pgf}};

		\node[anchor=north west,inner sep=0pt] (g) at (12.4,0) {\input{build/figures/si_emul_connectivity.pgf}};
		\node[anchor=north west,inner sep=0pt] (h) at (12.4,-4.1) {\input{build/figures/si_sim_offset_connectivity.pgf}};
		\node[anchor=north west,inner sep=0pt] (i) at (12.4,-8.2) {\input{build/figures/si_sim_connectivity.pgf}};

		\node[anchor=north west,inner sep=0pt] (j) at (16.5,0) {\input{build/figures/si_emul_average_weight.pgf}};
		\node[anchor=north west,inner sep=0pt] (k) at (16.5,-4.1) {\input{build/figures/si_sim_offset_average_weight.pgf}};
		\node[anchor=north west,inner sep=0pt] (l) at (16.5,-8.2) {\input{build/figures/si_sim_average_weight.pgf}};

		\node[rotate=90] (label0) at ($(j.east) + (0.7, 0.0)$) {\normalsize\bfseries Emulation};
		\node[rotate=90] (label1) at ($(k.east) + (0.7, 0.0)$) {\normalsize\bfseries Simulation (offset)};
		\node[rotate=90] (label2) at ($(l.east) + (0.7, 0.0)$) {\normalsize\bfseries Simulation};

		\node at ($(a.north west) + (0.2,-0.2)$) {\textbf{A}};
		\node at ($(b.north west) + (0.2,-0.2)$) {\textbf{B}};
		\node at ($(c.north west) + (0.2,-0.2)$) {\textbf{C}};
		\node at ($(d.north west) + (0.2,-0.2)$) {\textbf{D}};
		\node at ($(e.north west) + (0.2,-0.2)$) {\textbf{E}};
		\node at ($(f.north west) + (0.2,-0.2)$) {\textbf{F}};
		\node at ($(g.north west) + (0.2,-0.2)$) {\textbf{G}};
		\node at ($(h.north west) + (0.2,-0.2)$) {\textbf{H}};
		\node at ($(i.north west) + (0.2,-0.2)$) {\textbf{I}};
		\node at ($(j.north west) + (0.2,-0.2)$) {\textbf{J}};
		\node at ($(k.north west) + (0.2,-0.2)$) {\textbf{K}};
		\node at ($(l.north west) + (0.2,-0.2)$) {\textbf{L}};

		\begin{pgfonlayer}{background}
			\node[inner sep=6pt,rounded corners,fill=nicegreen!20,fit=(a) (label0)] (emulation) {};
			\node[inner sep=6pt,rounded corners,fill=nicered!20,fit=(b) (label1)] (simulation_offset) {};
			\node[inner sep=6pt,rounded corners,fill=niceblue!20,fit=(c) (label2)] (simulation) {};
		\end{pgfonlayer}
	\end{tikzpicture}
}
	\caption{%
		\textbf{Software simulations validate the neuromorphic emulation on BrainScaleS-2.}
		The panels \textbf{(A)}, \textbf{(D)}, \textbf{(G)} and \textbf{(J)} show the emulation results already presented in \cref{fig:chip} in the main part of this manuscript.
		The panels \textbf{(B)}, \textbf{(E)}, \textbf{(H)} and \textbf{(K)} illustrate the results of corresponding software simulations incorporating the amplitude offset within the synaptic currents present on the neuromorphic system, whereas \textbf{(C)}, \textbf{(F)}, \textbf{(I)} and \textbf{(L)} depict the simulation data acquired for ideal dynamics and hence without offset.
	}
	\label{fig:simulation_comparison_homeostasis}
\end{figure*}

\section{Validation of the analog emulation \label[supp]{sec:validation}}

Equivalent software simulations validate the main results on bistability in \gls{ei} networks of \gls{lif} neurons emulated on the analog BrainScaleS-2 system.
In the following, we very the emulation by first comparing to an accurate software simulation incorporating the offset in the synaptic current shown in \cref{fig:parametrization} G and H.
Most importantly, our results do not depend on the latter and persist for ideal synaptic dynamics.

The firing rates (\cref{fig:simulation_comparison_homeostasis} A, B and C), the histograms of synaptic weights (\cref{fig:simulation_comparison_homeostasis} D, E and F), the effective connectivity (\cref{fig:simulation_comparison_homeostasis} G, H and I) as well as the average synaptic weights (\cref{fig:simulation_comparison_homeostasis} J, K and L) are comparable for emulation and both simulation conditions.
However, for high input strengths, the firing rate $\nu$ of simulated networks exceeds the target rate.
Likewise, the effective connectivity and the average synaptic weights are slightly higher on the BrainScaleS-2 chip compared to the simulation.
This hints towards an overestimation of the \gls{psp} height on hardware which is most likely due to saturation effects promoted by the analog implementation.

\begin{figure*}[ht]
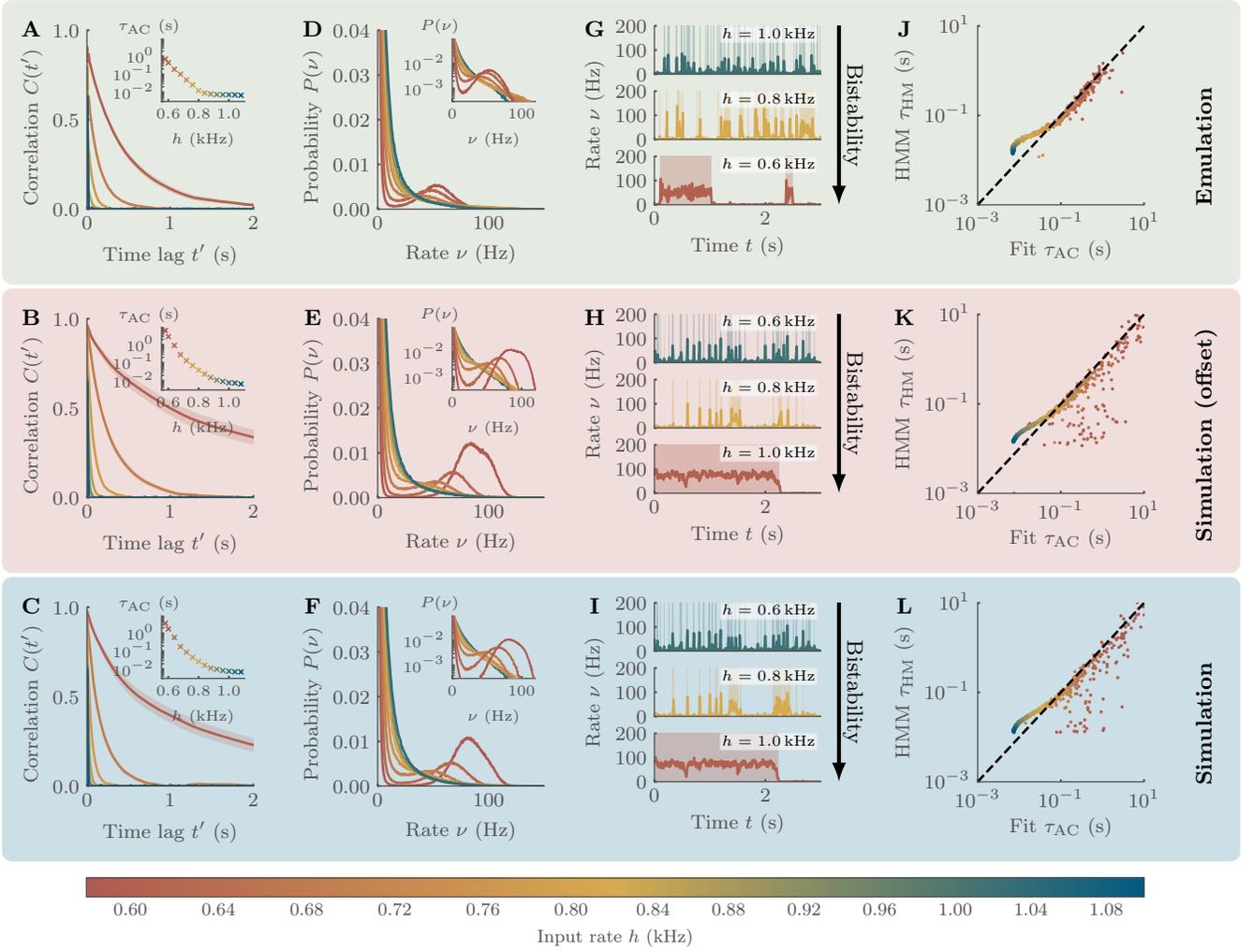

	\centering
	\begin{tikzpicture}
		\node[anchor=north west,inner sep=0pt] (a) at (0,0) {\input{build/figures/si_emul_ac.pgf}};
		\node[anchor=north west,inner sep=0pt] (b) at (0,-4.1) {\input{build/figures/si_sim_offset_ac.pgf}};
		\node[anchor=north west,inner sep=0pt] (c) at (0,-8.2) {\input{build/figures/si_sim_ac.pgf}};

		\node[anchor=north west,inner sep=0pt] (d) at (4.0,0) {\input{build/figures/si_emul_activity_distribution.pgf}};
		\node[anchor=north west,inner sep=0pt] (e) at (4.0,-4.1) {\input{build/figures/si_sim_offset_activity_distribution.pgf}};
		\node[anchor=north west,inner sep=0pt] (f) at (4.0,-8.2) {\input{build/figures/si_sim_activity_distribution.pgf}};

		\node[anchor=north west,inner sep=0pt] (g) at (8.0,0) {\input{build/figures/si_emul_activity.pgf}};
		\node[anchor=north west,inner sep=0pt] (h) at (8.0,-4.1) {\input{build/figures/si_sim_offset_activity.pgf}};
		\node[anchor=north west,inner sep=0pt] (i) at (8.0,-8.2) {\input{build/figures/si_sim_activity.pgf}};

		\node[anchor=north west,inner sep=0pt] (j) at (12.4,0) {\input{build/figures/si_emul_timescale_comparison.pgf}};
		\node[anchor=north west,inner sep=0pt] (k) at (12.4,-4.1) {\input{build/figures/si_sim_offset_timescale_comparison.pgf}};
		\node[anchor=north west,inner sep=0pt] (l) at (12.4,-8.2) {\input{build/figures/si_sim_timescale_comparison.pgf}};

		\scalebox{0.905}{\node[anchor=north west,inner sep=0pt] (x) at (1.03,-13.5) {\import{build/figures/}{colorbar.pgf}};}

		\draw[-latex,ultra thick] ($(g.north east) + (0.2,-0.15)$) -- ($(g.south east) + (0.2,0.8)$) node[midway,xshift=0.2cm,rotate=-90] {Bistability};
		\draw[-latex,ultra thick] ($(h.north east) + (0.2,-0.15)$) -- ($(h.south east) + (0.2,0.8)$) node[midway,xshift=0.2cm,rotate=-90] {Bistability};
		\draw[-latex,ultra thick] ($(i.north east) + (0.2,-0.15)$) -- ($(i.south east) + (0.2,0.8)$) node[midway,xshift=0.2cm,rotate=-90] {Bistability};

		\node[rotate=90] (label0) at ($(j.east) + (0.6, -0.1)$) {\vphantom{(}\small\bfseries Emulation};
		\node[rotate=90] (label1) at ($(k.east) + (0.6, -0.1)$) {\vphantom{(}\small\bfseries Simulation (offset)};
		\node[rotate=90] (label2) at ($(l.east) + (0.6, -0.1)$) {\vphantom{(}\small\bfseries Simulation};

		\node at ($(a.north west) + (0.2,-0.2)$) {\textbf{A}};
		\node at ($(b.north west) + (0.2,-0.2)$) {\textbf{B}};
		\node at ($(c.north west) + (0.2,-0.2)$) {\textbf{C}};
		\node at ($(d.north west) + (0.2,-0.2)$) {\textbf{D}};
		\node at ($(e.north west) + (0.2,-0.2)$) {\textbf{E}};
		\node at ($(f.north west) + (0.2,-0.2)$) {\textbf{F}};
		\node at ($(g.north west) + (0.2,-0.2)$) {\textbf{G}};
		\node at ($(h.north west) + (0.2,-0.2)$) {\textbf{H}};
		\node at ($(i.north west) + (0.2,-0.2)$) {\textbf{I}};
		\node at ($(j.north west) + (0.2,-0.2)$) {\textbf{J}};
		\node at ($(k.north west) + (0.2,-0.2)$) {\textbf{K}};
		\node at ($(l.north west) + (0.2,-0.2)$) {\textbf{L}};

		\begin{pgfonlayer}{background}
			\node[inner sep=6pt,rounded corners,fill=nicegreen!20,fit=(a) (label0)] (emulation) {};
			\node[inner sep=6pt,rounded corners,fill=nicered!20,fit=(b) (label1)] (simulation_offset) {};
			\node[inner sep=6pt,rounded corners,fill=niceblue!20,fit=(c) (label2)] (simulation) {};
		\end{pgfonlayer}
	\end{tikzpicture}
	\caption{%
		\textbf{Equivalent software simulations validate the emergence of bistable population activity for low inputs strengths and lead to comparable timescales.}
		The panels \textbf{(A)}, \textbf{(D)}, (\textbf{G}) and \textbf{(J)} show the emulation results already presented in \cref{fig:time_scale} in the main part of this manuscript.
		The panels \textbf{(B)}, \textbf{(E)}, \textbf{(H)} and \textbf{(K)} illustrate the results of corresponding software simulations incorporating the amplitude offset within the synaptic currents present on the neuromorphic system, whereas \textbf{(C)}, \textbf{(F)}, \textbf{(I)} and \textbf{(L)} depict the simulation data acquired for ideal dynamics and hence without offset.
	}
	\label{fig:simulation_comparison}
\end{figure*}

The bistable population activity and the associated emerging autocorrelation persist in simulation.
More specifically, the autocorrelation functions as well as the estimated autocorrelation times of simulation and emulation mostly coincide (\cref{fig:simulation_comparison} A, B and C), the distribution of the population activity shows a similar bimodal trend (\cref{fig:simulation_comparison} D to I) and the autocorrelation times are well described by the \gls{hmm} (\cref{fig:simulation_comparison} J, K and L).
Hence, we conclude that the bistable behaviour as well as the associated emergent autocorrelations are indeed a result of the model dynamics.

\begin{figure*}[ht]
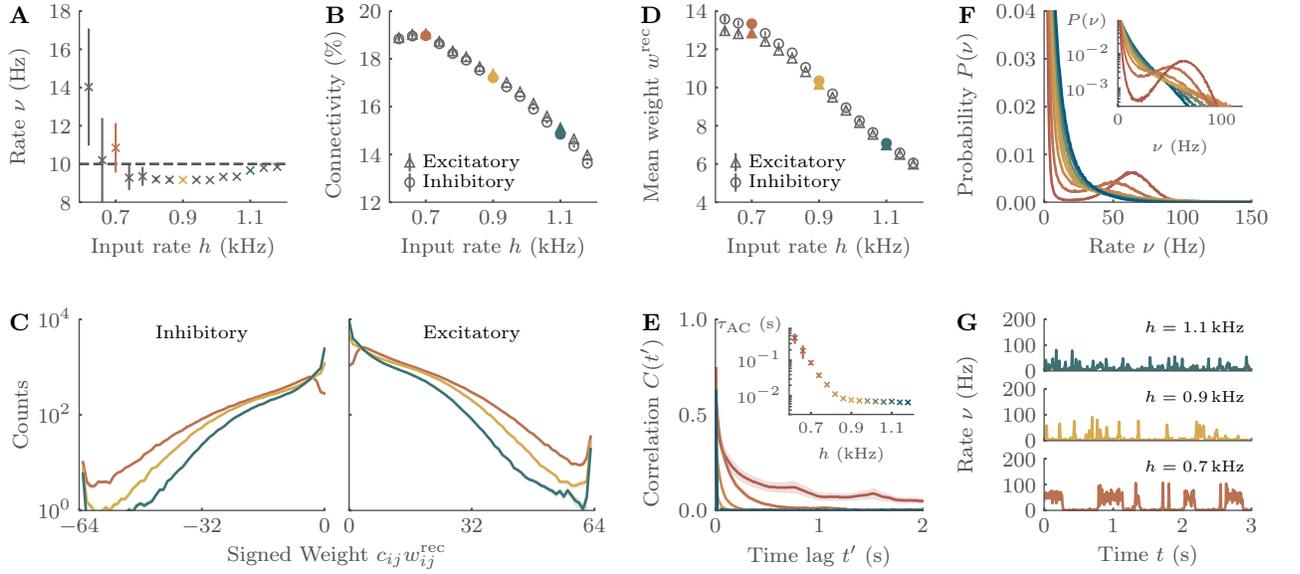

	\centering
	\begin{tikzpicture}
		\node[anchor=north west,inner sep=0pt] (a) at (0,0) {\input{build/figures/si_hom_final_rate.pgf}};
		\node[anchor=north west,inner sep=0pt] (b) at (4.2,0) {\input{build/figures/si_hom_connectivity.pgf}};
		\node[anchor=north west,inner sep=0pt] (c) at (0,-4.1) {\input{build/figures/si_hom_weight_histogram.pgf}};
		\node[anchor=north west,inner sep=0pt] (d) at (8.4,0) {\input{build/figures/si_hom_average_weight.pgf}};
		\node[anchor=north west,inner sep=0pt] (e) at (8.4,-4.1) {\input{build/figures/si_hom_ac.pgf}};
		\node[anchor=north west,inner sep=0pt] (f) at (12.6,0) {\input{build/figures/si_hom_activity_distribution.pgf}};
		\node[anchor=north west,inner sep=0pt] (g) at (12.6,-4.1) {\input{build/figures/si_hom_activity.pgf}};
		\node at ($(a.north west) + (0.2,-0.2)$) {\textbf{A}};
		\node at ($(b.north west) + (0.2,-0.2)$) {\textbf{B}};
		\node at ($(c.north west) + (0.2,-0.2)$) {\textbf{C}};
		\node at ($(d.north west) + (0.2,-0.2)$) {\textbf{D}};
		\node at ($(e.north west) + (0.2,-0.2)$) {\textbf{E}};
		\node at ($(f.north west) + (0.2,-0.2)$) {\textbf{F}};
		\node at ($(g.north west) + (0.2,-0.2)$) {\textbf{G}};

	\end{tikzpicture}
	\caption{%
		\textbf{A second implementation of homeostatic regulation leads to similar results of bistable population activity on BrainScaleS-2.}
		Homeostatic regulation incorporating stochastic rounding promotes bistable population activity for low input rates $h$.
		\textbf{(A)} This choice of regulation likewise adjusts the population rate $\nu$ close to a target value (dashed line).
		\textbf{(B)} The connectivity slightly exceeds the homeostatic regulation with probabilistic update acceptance.
		\textbf{(C)} The distribution of weights $c_{ij}w_{ij}^\mathrm{rec}$ is comparable to the one reached with stochastic update acceptance.
		\textbf{(D)} However, the average weight $w^\mathrm{rec}$ is slightly lower.
		\textbf{(E)} The population activity $\nu$ features exponentially shaped autocorrelation functions $C(t')$ with autocorrelation times $\tau_\mathrm{AC}$ of same magnitude as the ones presented in main manuscript.
		\textbf{(F)} The distribution of the population activity $P(\nu)$ again resembles a bimodal trend for decreasing values of $h$, explaining the emergent autocorrelation times.
		\textbf{(G)} This bimodal trend is reflected in form of phases of high and low activity for low $h$.
	}
	\label{fig:hom_comparison}
\end{figure*}

\section{Comparison to a different implementation of homeostatic regulation \label[supp]{sec:hom_comparison}}

The neuromorphic implementation puts constraints on the range of implementable homeostatic regulation formulas.
As indicated by \cref{fig:stability}, the limited range, integer arithmetic of the \gls{ppu} requires additional mechanisms -- like the stochastic acceptance of weight updates -- to guarantee smooth convergence of the synaptic weights and in turn the attainment of the desired target rate.
While the stochastic acceptance renders the resulting weight distributions diverse (\cref{fig:chip}D), it likewise assigns many synapses a weight of zero (\cref{fig:chip}F).
To investigate whether the resulting bistable population activity is a result of our specific choice of homeostatic regulation, we implemented a second mechanism which aims to adapt the afferent synaptic weights of each neuron more uniformly.
Specifically, we utilize the same form as the one presented in \cref{eq:rule}, but with a smaller learning rate $\lambda = 0.39$ and instead of a stochastic acceptance probability added a small uniformly drawn random number to each update to realize stochastic rounding.
This mechanism likewise regulates the network activity (\cref{fig:hom_comparison} A to D), and, moreover, reveals emerging autocorrelations for low $h$ of similar magnitude (\cref{fig:hom_comparison}).
Different to the rule presented in the main manuscript, the connectivity increases monotonously with decreasing $h$ and only slightly decreases for very small $h$ again (\cref{fig:hom_comparison}B).
Again, these correlations stem from phases of low and high activity (\cref{fig:hom_comparison} F and G) in direct accordance with our previous results.

\section{Solution of the mean-field Langevin equation}

For the mean activity of a finite network, we obtain to leading order in system size
\begin{equation}\label{si_eq:MF_noise}
    \dot{\rho}(t) = h - a\rho(t) -b\rho^2(t) + \sigma\sqrt{\rho(t)/N}\eta(t)\, ,
\end{equation}
where $\eta(t)$ is Gaussian white noise with mean zero and unit variance and $\sigma/\sqrt{N}$ is a system-size dependent factor.
This (Ito) Langevin equation is equivalent to the Fokker Planck equation
\begin{align}\label{si_eq:FP_noise}
    \dot{P}(\rho,t) = &-\frac{\partial}{\partial\rho}\left[\left(h-a\rho(t)-b\rho^2(t)\right)P(\rho,t)\right]\nonumber\\
                    &+\frac{\sigma^2}{2N}\frac{\partial^2}{\partial\rho^2}\left[\rho(t)P(\rho,t)\right]\, .
\end{align}
Imposing $\dot{P}(A,t)=0$ and simplifying $f(\rho)=h-a\rho-b\rho^2$ as well as $g(\rho)=\rho\sigma^2/2N$, we get an equation for the stationary state $P^\ast(\rho)$
\begin{align}
    0
    &= -\frac{d}{d\rho}\left[f(\rho)P(\rho)\right] + \frac{d^2}{d\rho^2}\left[g(\rho)P(\rho)\right]\\
    &= \frac{d}{d\rho}\left(-f(\rho)P(\rho) +  \frac{d}{d\rho}\left[g(\rho)P(\rho)\right]\right) = \frac{d}{d\rho} j,
\end{align}
where $j$ reminds of the current in a continuity equation that has to be constant.
Imposing the zero-flux condition ($j=0$) one can derive the steady state probability distribution as usual~\cite{gardiner_handbook_1985}.
For this, we have to solve
\begin{align}
 \frac{d}{d\rho}\left[g(\rho)P(\rho)\right] = f(\rho)P(\rho).
\end{align}
Replacing $\Gamma(\rho)=g(\rho)P(\rho)$, we have to solve
\begin{align}
    \frac{d}{d\rho}\Gamma(\rho) = \Gamma(\rho)f(\rho)/g(\rho),
\end{align}
and obtain
\begin{align}
    \Gamma(\rho) = e^{\int \frac{f(\rho)}{g(\rho)}d\rho}.
\end{align}
In our case, $f(\rho)/g(\rho)= \left(h/\rho -a -b\rho\right)2N/\sigma^2$ such that
\begin{align}
    P(\rho) = \Gamma(\rho)/g(\rho) \propto \frac{2N}{\rho\sigma^2}e^{\frac{2N}{\sigma^2}\left(h\ln\rho-a\rho-\frac{b}{2}\rho^2\right)},
\end{align}
Rewriting $1/\rho=e^{-\ln{\rho}}$, we finally arrive at
\begin{equation}
    P(\rho) = \mathcal{N}\exp\left\{-\frac{2N}{\sigma^2}V(\rho)\right\}\, ,
\end{equation}
with a normalization constant $\mathcal{N}$ and the potential
\begin{align}
    V(\rho) = \left(\frac{\sigma^2}{2N}-h\right)\ln\rho + a\rho + \frac{b}{2}\rho^2\, .
\end{align}